\documentclass[galley, draft, jgrga]{AGUTeX}

\usepackage{url}
\usepackage{pdflscape}

 \usepackage[dvips]{graphicx}
 \usepackage[toc,page]{appendix}

  \setkeys{Gin}{draft=false}

\authorrunninghead{XIE ET AL.}

\titlerunninghead{Energy-dependent Electron and Proton Onset Times}

\authoraddr{H. Xie,
Department of Physics, The Catholic University of
America, 200 Hannan Hall, Washington, DC 20064, USA.
(hong.xie@nasa.gov)}

\begin{document}

%% ------------------------------------------------------------------------ 
%%
%
%  TITLE
%
%% ------------------------------------------------------------------------ 
%%

\title{Energy Dependence of SEP Electron and Proton Onset Times}

%% ------------------------------------------------------------------------ 
%%
%
%  AUTHORS AND AFFILIATIONS
%
%% ------------------------------------------------------------------------ 
%%

%Use \author{\altaffilmark{}} and \altaffiltext{}

% \altaffilmark will produce footnote;
% matching altaffiltext will appear at bottom of page.
% May use \\ to start a new line.

\authors{H. Xie, \altaffilmark{1,2} P. M\"akel\"a,\altaffilmark{1,2} N. Gopalswamy, \altaffilmark{2} and  O. C. St. Cyr,\altaffilmark{2}}

\altaffiltext{1}{Department of Physics, The Catholic University of America, 
Washington DC, USA.}
\altaffiltext{2}{Code 670, NASA/Goddard Space Flight Center, Greenbelt, Maryland, 
USA}

%% ------------------------------------------------------------------------ 
%%
%
%  ABSTRACT
%
%% ------------------------------------------------------------------------ 
%%

% >> Do NOT include any \begin...\end commands within
% >> the body of the abstract.

\begin{abstract}
We study the large solar energetic particle (SEP) events that were detected by GOES in the $>$ 10 MeV energy channel during December 2006 to March 2014. 
%Using observations from STEREO A, B and SOHO, we %are able to determine accurately the solar release time of SEP electrons and protons. We first compute connection angles (CA) between the solar sources of the events and the footpoints of the spiral magnetic field lines for each spacecraft. %By choosing the smallest CA among the three spacecraft, 
We derive and compare solar particle release (SPR) times for the 0.25--10.4 MeV electrons and 10--100 MeV protons for the 28 SEP events.
In the study, the electron SPR times are derived with the time-shifting analysis (TSA) and the proton SPR times are derived using both the TSA and the velocity dispersion analysis (VDA).
Electron anisotropies are computed to evaluate the amount of scattering for the events under study.
Our main results include: 1) near-relativistic electrons and high-energy protons are released at the same time within 8 min for most (16 of 23) SEP events. 
2)There exists a good correlation between electron and proton acceleration, peak intensity and intensity time profiles. 
%2) The first arriving electrons propagate nearly scatter-free for 23 of 27 SEP events.
3) The TSA SPR times for 90.5 MeV and 57.4 MeV protons have maximum errors of 6 min and 10 min compared to the proton VDA release times, respectively,
while the maximum error for 15.4 MeV protons can reach to 32 min.
4) For 7 low-intensity events of the 23, large delays occurred between 6.5 MeV electrons and 90.5 MeV protons relative to 0.5 MeV electrons.
Whether these delays are due to times needed for the evolving shock to be strengthened or due to particle transport effects remains unsolved.
\end{abstract}

%% ------------------------------------------------------------------------ 
%%
%
%  BEGIN ARTICLE
%
%% ------------------------------------------------------------------------ 
%%

% The body of the article must start with a \begin{article} command
%
% \end{article} must follow the references section, before the figures
%  and tables.

\begin{article}

%% ------------------------------------------------------------------------ 
%%
%
%  TEXT
%
%% ------------------------------------------------------------------------ 
%%

\section{Introduction}
The origin of energetic particles accelerated in solar events is still an open question. While flares and shocks driven by coronal mass ejections (CMEs) are believed to be two sources of solar energetic particle (SEP) acceleration in impulsive and gradual SEP events respectively \citep[e.g.][]{Reames1999}, it is not clear what the exact flare-related acceleration mechanism in the impulsive SEP events is or where the CME-driven shocks most efficiently accelerate particles and when the particles are released in gradual SEP events.

Electron release has been observed to temporally coincide with type III radio bursts at the Sun and traveling along open field lines into interplanetary space \citep[see][]{Lin1985}.
Using observations of the Three-dimensional Plasma and Energetic Particles instrument
\citep[3DP;][]{Lin1995} on the Wind spacecraft, \citet{Wang2006} studied three electron events and found two distinct injections of electrons: that of low-energy electrons at energies $\sim$ 0.4 to 6--9 keV began 9.1 min before the type III radio burst and that of $\sim$ 13 to 300 keV electrons started 7.6 min after the type III burst. 
%They suggested that the low-energy electrons were likely to be the source of the type III radio bursts, and the delayed high-energy injection occurred when the associated CME reached heights of 1--6 Rs.
Delays of 10 min up to half an hour between electron release time at the Sun and solar electromagnetic emissions (EM) have been reported by other works \citep[e.g.][]{Cliver82,Kallenrode91,Krucker1999, Haggerty02}. 
\citet{Krucker1999} showed evidence that some electron events are not related to type III bursts. 
%These electron events can have a delayed release up to half an hour after the type III burst.
They found that the electron events appeared to be related to  the passage of large-scale coronal transient waves, also called EIT waves or Extreme Ultraviolet (EUV) waves \citep{Thompson1998, Thompson2000}, over the footpoint of the field line connected to the spacecraft.
Although the nature of EUV waves is still largely debated,
past studies  using low-cadence ultraviolet images ($>$12 minutes)
showed that  EUV waves are correlated with CMEs rather than flares \citep{Plunkett1998, Cliver1999}. 
Based on more recent three-dimensional stereoscopic analyses, the EUV waves are generally believed to be the imprint of the CME driven shock on solar surface  \citep[e.g.][]{Veronig2008, Patsourakos2009}.

Proton release is probably more complicated than electron release. \citet{Krucker2000} studied the timing of proton onsets in the energy range from 30 keV to 6 MeV. They found that the release of the protons appears to be energy-dependent. The most energetic protons are possibly released simultaneously with the electrons while lower-energy protons are released $\sim$ 0.5 to 2 hrs later than electrons. They also found that protons with energies between 0.03 and 6 MeV are released high in the corona, around 1--10 Rs above the electrons.  Their results are consistent with studies by \citet{Kahler1994} and \citet{Gopalswamy2012} on the CME heights at the time of SEP release. \citet{Kahler1994} analyzed $>$ 10 MeV proton events and found that the peak of the intensity profile for $>$ 10 MeV protons occurs when the associated CME reaches heights of 5--15 Rs. \citet{Gopalswamy2012} examined the onset times and release heights of energetic particles using the ground-level enhancement (GLE) events. They found an earlier release time and a lower release height of CMEs for this highly energetic subset of events.

Although both SEP solar particle release (SPR) times and EM onsets have been discussed at length in the past, comparison between electron  release times and proton release times has been discussed only in a few papers \citep[e.g.][]{Cliver82,Haggerty09,Koulou15,Kahler2003,Posner07}.
In \citet{ Posner07}'s study, the author adopted the prevailing assumption of simultaneous release of electrons and protons, but he also pointed out that ``release of protons before electrons (and vice versa) is possible [E. Roelof, D. Haggerty, personal communication, 2006]''. 
Using a group of 32 historic GLE events, \citet{Cliver82} found that delays of 10 minutes between 100 keV
and 1 MeV electron SPRs and $\le$ 5 minute delays between 2 GeV proton and 1 MeV electron SPRs. 
Delays of 10 to 50 minutes in the proton SPRs relative to metric type II onsets for well connected events were found in the smaller GLE events.
\citet{Kahler2003} compared the onset of relativistic electrons and protons of GLEs from solar cycle 23. 
They found that half of GLE events the relativistic proton injection preceded that of electrons,
however, the low intensity GLEs tend to have a later time for the proton injection. 
Recently, \citet{Koulou15} compared the proton and electron release as inferred from VDA based on 
Wind/3DP and ERNE data, and found a 7-min average dalay of near-relativistic electrons with respect to deka-MeV protons.
\citet{Haggerty09} studied 19 electron beam events using EPAM 38-315 keV data, and found that for 11 of the 19 events the arrival of 50-100 MeV protons followed by electrons within $\sim$ 3 min. On the other hand, the remaining 8 events show a broad 5-25 minute delays of the protons relactive to the electron injections.

%We use 20MeV proton intensities from the EPACT instrument on Wind
%\citet{Kahler1994} analyzed greater than 10 MeV proton events and found that the peak of the intensity profile for $>$ 10 MeV protons occurs when the associated CME reaches heights of 5--15 Rs. \citet{Gopalswamy2012} examined the onset times and release heights of energetic particles using the ground-level enhancement (GLE) events. They found an earlier release time and a lower release height of CMEs for this highly energetic subset of events. 

In this paper, we study large SEP events with a peak $>$ 10 MeV proton flux above 10 $cm^2 sr^{-1} s^{-1}$ as observed by GOES from October 2006 (the launch of the Solar and Terrestrial Relations Observatory (STEREO))  to March 2014.
The proton SPR times at various energies from 10 MeV to 131 MeV are investigated and compared to the release times of 0.25 MeV--10.4 MeV electrons and solar EM onsets. %including type II and type III radio emission onsets.
The exact time when energetic particles are first released at the Sun is crucial to understanding the particle acceleration and where it takes place.
This is the first systematic study and comparison between  electron and proton SPRs for large SEP events in the new STEREO era.
The paper aims to address the following key issues:
1) Are protons and electrons accelerated by the same source and released simultaneously at the Sun?
%2) Where is the acceleration source location for SEP particles and what is their acceleration mechanism?
2) What is the acceleration time needed for protons and electrons to reach high energies and are the acceleration times energy dependent?

\section{Observations and Data Analysis}

\subsection{Event Selection}

From October 2006 to March 2014, GOES have observed 35 large SEP events with peak intensity great than 10 $cm^2 sr^{-1} s^{-1}$ in the $>$ 10 MeV proton channel (\url{http://cdaw.gsfc.nasa.gov/CME_list/sepe/}).
In this paper we selected 28 of the 35 SEP events 
%with the GOES $>$ 10 MeV proton flux above 10 cm$^{-2}$ s$^{-1}$ sr$^{-1}$ 
and excluded 6 events where the Energetic and Relativistic Nuclei and Electron
instrument (ERNE) measurements have a data gap 
and 1 event where only a mild flux enhancement ($< 10\%$) was seen above the background level.   
When an SEP event has multiple-increases of flux we define the earliest rise as its onset time and count it as one event.
We divide these 28 events into two groups: SOHO (the Solar and Heliospheric Observatory) SEP events and STEREO SEP events.
By choosing the smallest connection angles (CA) between SEP solar locations and magnetic foot-points of each spacecraft, we define whether a SEP event is SOHO SEP or STEREO SEP.
We compute the longitude of connection footpoint by assuming Parker spiral theory:
\begin{equation}
\phi_0 = D\Omega/V_{slw} + \phi , 
\end{equation}
where $\phi$ and $\phi_0$ are the spacecraft longitude and its solar connection footpoint longitude, $D$ is the distance to the Sun, $V_{slw}$ is the average in-situ solar wind speed observed by the spacecraft, and $\Omega$ is the solar rotation rate based on a Sidereal rotation period of 24.47 days. 
CA is then given by:
\begin{equation}
CA = \phi_0 -\phi_{src} ,
\end{equation}
where $\phi_{src}$ is the solar source longitude of SEPs. It should be noted that an average spread of $\sim 30^{\circ}$ 
between active regions and  source surface connection footpoints has been reported in previous statistical studies  \citep[e.g.][]{Nitta06,Wiedenbeck2013} 
and an uncertainty of CA angles as large as $20^\circ$ was found using various methods in \citet{Lario14}.
%Thus even when a SEP CA is $60^{\circ}$ we may call the SEP event a magnetically well-connected event. 

The solar source locations were identified as the locations of associated flares or eruptive prominences in movies
of EUV images by the Atmospheric Imaging Assembly \citep[AIA;][]{Lemen2012} on the \emph{Solar Dynamics
Observatory} (SDO) spacecraft and by the EUV Imager \citep[EUVI;][]{Wuelser2004,Howard2008} on the STEREO spacecraft.
Additional information on the events has been extracted
from the GOES flare list (\url{http://www.lmsal.com/solarsoft/latest_events/}), the CDAW CME catalog (\url{http://cdaw.gsfc.nasa.gov/CME_list}),
and the type II radio burst lists compiled by the Wind and STEREO data center (\url{http://ssed.gsfc.nasa.gov/waves/data_products.html}),
and type III radio burst data (\url{http://cdaw.gsfc.nasa.gov/images/wind/waves/}).

%(\url{http://soho.nascom.nasa.gov/data/ancillary/attitude/roll/nominal_roll_attitude.dat})

In-situ observations including the Electron Proton and Helium Instrument \citep[EPHIN;][]{Mellin1995} and ERNE  \citep{Torsti1995} on the SOHO spacecraft, 
and the High Energy Telescope \citep[HET;][]{von2008}, Low Energy Telescope \citep[LET;][]{Mewaldt2008}, and the Solar Electron Proton Telescope \citep[SEPT;][]{Mellin2008} on the STEREO spacecraft are used for the determination of the SEP onsets.

SOHO/ERNE covers the energy range from 1.58 to 131 MeV of protons by using two different sensors. The Low-Energy Detector (LED) operates in the range 1.58 MeV to 12.7 MeV and the High-Energy Detector (HED) from 13.8 MeV to 131 MeV. 
We determined proton onset times of SOHO SEP events using SOHO/ERNE 1-minute averages in proton energy channels from 13.8 MeV to 131 MeV.
We used only HED channels 
since the small geometric factor at the lowest energies yields a relatively high intensity
at 1-count level, which make it difficult to determine the event onset time accurately \citep{Vainio13}. 
%lowest energy channels of ERNE LED tend to have large background levels which can cause large errors in onset time determination in many events \citep{Laitinen2010}. Also, LED has low statistics at the highest-energy channels due to its small geometric factor (about 1 $cm^2 sr$) \citep{Vainio13}. 
We used SOHO/EPHIN 1-minute averages in energy channels from 0.25 to 10.4 MeV to determine electron onset times. 
When EPHIN data were not available, we used instead 1-minute averages of the 230--392 keV electron data from Wind/3DP or the 175--315 keV electron data from the Electron, Proton, and Alpha Monitor \citep[EPAM;][]{Gold1998} on the \emph{Advanced Composition Explorer} (ACE).
% define the onset as the point where the flux is above  by there standard deviations above background.
For STEREO SEP events,  we used STEREO/SEPT 1-minute averages (sun-direction) in the electron energy channel 0.255--0.295 MeV, STEREO/HET 1-minute averages in the electron channel 2.8--4 MeV and proton energy channels from 40 to 100 MeV, and LET 1-minute averages in proton energy channel 10--12  MeV to determine the onset times of electrons and protons.
We did not use HET data in the proton channel 13.6--15.1 MeV due to its large data gaps in many events. 
Instead LET standard data in the proton energy channel 10--12  MeV was used.

To estimate the scattering effect of the first arriving particles, we compute the anisotropy of the electrons using Wind/3DP data for the SOHO events and SEPT data for the STEREO events. 
The anisotropy of the protons was not considered in this study due to the lack of data available in instruments including ERNE, EPHIN, and HET, instead we use the
VDA to evaulate their scattering effects.
Furthermore, note that SOHO rotates 180\deg every three months and the pointing direction of ERNE and EPHIN will change from being sunward along the nominal Parker spiral direction to being perpendicular to that. Thus it is likely that both ERNE and EPHIN will miss the first arriving particles when SOHO's roll angle is 180\deg. We estimate the uncertainty by comparing the EPHIN and ERNE data with the available Wind/3DP data or ACE/EPAM data.
%0.25--0.70 MeV channel with ACE/EPAM 0.175--0.315 MeV data, a maximum of $\sim$ 10 min error has been found.

Finally, SEP intensity measurements can suffer from contamination. Possible causes of contamination include 1) particle misidentification, e.g. the presence of electrons in the proton channels and 2) missing particle energy, e.g. high-energy protons (electrons) deposit only a fraction of the energy at the detector and thus are counted as low-energy protons (electrons). \citet{Posner07} analyzed extensively EPHIN electron and proton measurements and found that some contamination (see also \citet{delPeral01}) and instrument dead time problems exist during the main phase of the SEP event, but the onset time determination using EPHIN electron and proton measurements can be done reliably. \citet{Haggerty02, Haggerty03} used simulations to examine contamination in the ACE/EPAM electron channels and they concluded that while the effect can be significant in the lowest-energy channels (E'1 and E'2), it is negligible in the highest two channels E'3 (102--175 keV) and E'4 (175--312 keV).
Other contamination factors are X-rays which will increase COSTEP front detector count rate \citep{Posner07, Klassen05}. 
In this study, we have excluded contaminations that may caused 
%the first flux pulse in the apparent double-peak electron events caused
by the X-ray or other particle energy misidentifations.
 
%In this current study, only high-energy proton channels ($>$ 13 MeV) from ERNE and electron channels ($>$ 250 keV) from EPHIN and ACE/EPAM were used to determine the onset times of SEP events. 
So far, no simulations have been conducted to evaluate contamination in ERNE data (Valtonen, personal communication) or reported for STEREO HET, LET and SEPT data. Therefore, we used the SEP intensity data from their instrument websites as provided with no additional contamination corrections.

\subsection{SEP Onset Time Determination}

We used an intersection slope method to determine the onset times of first arriving particles.
We first make linear fits for the background and increasing logarithmic fluxes of SEPs respectively and then take the intersection of the two lines as the onset time.
The uncertainty of the intersection slope method was estimated by the earliest and latest possible onset times, which were determined by the intersection with the background level, $\pm 3\sigma/(slope-error_{slope})$, similar to the method used in \citet{Miteva14}.
Figure~\ref{Fig_2slp} shows the procedure for (a) the 2011 August 8 SEP event and (b) the 2011 August 4 SEP event. The SEP event in (a) has a rapid increase of flux, allowing for an accurate determination of the onset times while the SEP intensity in (b) has a slow rise which caused a larger uncertainty, as indicated by the gray rectangle in the figure.
The obtained uncertainties of the intersection slope method range from $\pm$ 2 min to 9 min for the 28 SEP events. A similar uncertainty of ten minutes has been reported in other studies using alternative methods
 \citep[e.g.][]{Huttunen05, Vainio13}.

In general, the uncertainty of the intersection slope method itself (caused by the background flux fluctuation) is relatively small when compared with the high background errors.
%If estimated uncertainties are less than 5 minute, we then set them as $\pm$ 5 minute.
%It is known that the background of SEP flux will affect the determination of SEP onset times \citep[e.g.][]{Laitinen2010, Lintunen2004,Saiz2005}. 
%due to the previous SEP event or instrumental noise

It is well known that if the background flux of an SEP event is too high, it will mask the real onset time of SEPs  \citep[e.g.][]{Lintunen2004, Laitinen2010}.
To illustrate the background effect on the onset time, we over-plotted two elevated onset levels (ratio of background level to peak flux) of  $\sim 5\%$ (orange dashed line) and $\sim 8\%$ (blue dashed line) in Figure~\ref{Fig_2slp} (b). They introduced an error of $\sim$ 23 min and $\sim$ 32 min, respectively. 
The error caused by the high background can be estimated by:
\begin{equation}
ERR_{bglv} = (Int_{ont1} - Int_{ont0})/slope_{fit}
\end{equation}

where $Int_{ont1}$ and $Int_{ont0}$ are the logarithm of the SEP intensity at onset level 1 and onset level 0 and $slope_{fit}$ is the linear fit slope to the logarithm of the SEP intensity. 
In this work, we used equation (3) to correct the background effect by choosing a normalized onset level as $1\%$ of maximum flux in all data set, where the maximum flux is defined as the SEP prompt peak within 6 hours of the onset.
Also, note that the data time averages set a lower limit to the onset uncertainty.  
For example, in Figure~\ref{Fig_2slp} (b) we applied a 3-point running smoothing average to the 1-minute intensity data to make the early rise of the event more easy to see, which 
set a lower limit of 3 min for this case.
The uncertainty listed in Table 1 has a lower limit of time averages and upper limit of the background errors and errors of contaminations.
Note that, the first small flux increase in Figure~\ref{Fig_2slp} (a) was caused by X-ray contamination, we have included an uncertainty of $\sim$ 12 min as a lower limit in Table 1.

\newcommand{\txw}{\textwidth}

   \begin{figure}
    \begin{center}
    \includegraphics[width=0.92\txw, height = 0.36\txw ]{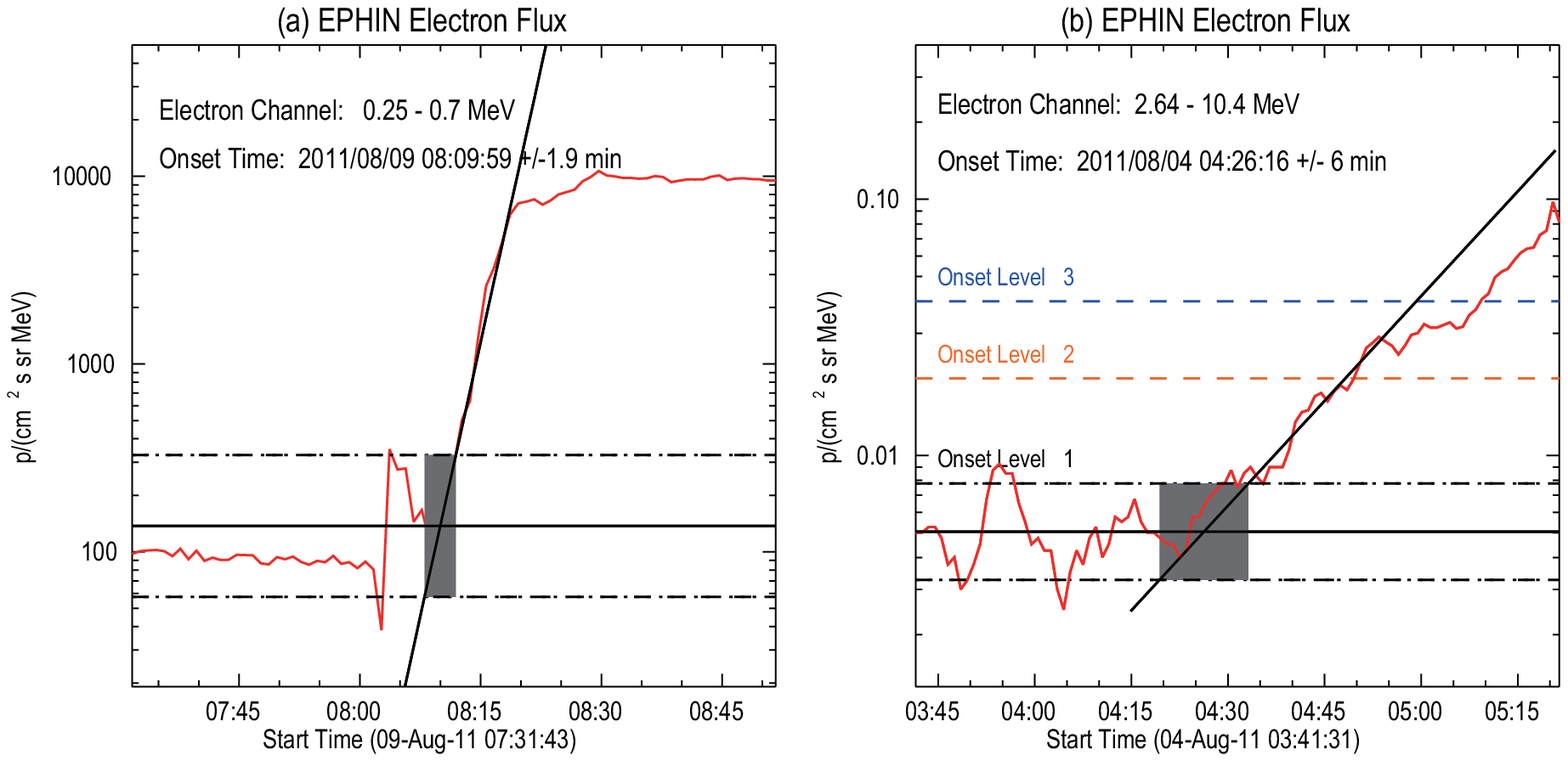}
    \end{center}
 \caption{1-minute averaged intensity of near-relativistic electrons (2.64--10.4 MeV) from SOHO/EPHIN observations for: a) the 2011 August 9 SEP and b) the 2011 August 4 SEP. Horizontal lines give the background level (solid), the average intensity during a pre-event time interval, and the background $\pm$ 3$\sigma$ levels (dashed--dotted). The inclined line is the linear fit to the logarithm of the SEP intensity during the early rise of the event. 
 The onset time is the time of intersection of this line with the background, the gray rectangle indicates the uncertainty. 
The orange and blue dashed lines in (b) are two assumed background levels, illustrating how the elevated background levels affect SEP onset times.}\label{Fig_2slp}
 \end{figure}

%    \includegraphics[width=0.92\txw]{new_Fig1.tif}
%       \mbox{[width=0.46\txw, height = 0.38\txw ]
%    \includegraphics[width=0.46\txw, height = 0.38\txw ]{new_Fig1b.eps}
%    }\par

%as defined by the times of intersection of the fitted profile with the background $\pm$ 3$\sigma$ levels.

%Comparing with other alternative methods, our method yield similar uncertainties as other methods commonly used in the determination of onset times (eg.). }

%Note that all SEP intensity measurements have contamination issue more or less. 

\subsection{Solar Particle Release Time Determination}
To infer the electron and proton release time at the Sun, time-shifting analysis (TSA) and velocity dispersion analysis (VDA) are two commonly used methods in the past works \citep[e.g.][]{Krucker1999,Tylka2003,Malandraki12,Vainio13}.
%Assuming that the first observed particles travel along the mean magnetic field scatter-free with zero pitch-angle approximation, 
The TSA computes the SPR time by shifting the onset time by particle traveling time along the nominal Parker spiral field line:  $t_{SPR} = t_{onset} - l/v$, where $l$ is the nominal path length from the Sun to the spacecraft and $v$ is the particle speed.
The nominal path length is computed using the Parker spiral field-line model with the average solar wind speed measured in-situ at the observing spacecraft.
The result of TSA represents a latest possible release of SEP particles.
It is a good approximation if the SEP particles travel nearly scatter free at nearly zero pitch angle along the magnetic field line.
%Note that the SPR times inferred from TSA are the upper limit of the release of SEP particles. 
For those particles which experience strong scattering, the TSA method can introduce large errors, especially for the protons. The error of TSA SPRs is given by:
\begin{equation}
ERR_{tsa} = (l_{sct}-l_{nom})/v
\end{equation}
where $l_{sct}$ and $l_{nom}$ are scattering path length and nominal path length and $v$ is the particle speed. 
Note that the TSA method is a good approximation for near-relativistic electrons and the TSA error for electrons is relatively small due to their extremely high speeds.  
For example, considering a path length range of 1.25 AU to 2 AU, the uncertainty of the TSA is given by $dt = (2 AU - 1.25 AU )/v$.  
For 1 MeV and 0.25 MeV electrons and 100 MeV, 50 MeV and 10 MeV protons, the corresponding errors are $\sim$ 6 min, 8 min, 13 min, 19 min and 40 min.

%In this we used the TSA to estimate SEP SPR times of all the events in Table 1. The corresponding uncertainty has been listed in the table.

%it will take longer for them to travel from the Sun to the spacecraft, and their release times would be earlier than the TSA release times.

The velocity dispersion analysis (VDA) is another method commonly used to estimate the release time of SEPs and their travel path 
length. The VDA method is based on the assumption that particles at all energies are released simultaneously and travel the same path length \citep{Krucker1999, Tylka2003, Vainio13}.

The particle arrival time at 1 AU is given by:

\begin{equation}
t_{{onset}}(E) = t_0 + 8.33 \frac{{min}}{{AU}}L(E)\beta^{-1}(E)
%t_{\text{onset}}(E) = t_0 + 8.33 \frac{\text{min}}{\text{AU}}L(E)\beta^{-1}(E)
\end{equation}

where $t_{{onset}}(E)$ is the onset time in minutes observed in different energy E, $t_0$ is the
release time in minutes at the Sun, $L$ is the path length (AU) travelled by the particle and
$\beta^{-1}(E)= c/v(E)$ is the inverse speed of the particles.
If energetic particles travel the same path length and are released at the same time then a linear dispersion relation can be obtained by plotting particle onset times versus $\beta^{-1}$.
The slope and intersection of the linear fit yield the path length and the particle release time
at the Sun, respectively.

\section{Statistics and Analysis Results}

\subsection{Event Catalog}

Table~\ref{Table1} summarizes the timing of 28 selected SEP events and associated solar eruptions.
The first and second columns of the table list SEP event number and date.
The  numbers 1--17 denote 17 SOHO SEP events and the numbers S1--S11 denote 11 STEREO SEP events.
The third and fourth columns of the table show the release times of SEP electrons with uncertainty in parentheses.
e1 and e2 represent 0.25--0.7 MeV and 2.64--10.4 MeV electrons for SOHO SEP events
and 0.255--0.295 MeV and 2.8--4.0 MeV electrons for STEREO SEP events.
The fifth to seventh columns of the table show the release times of SEP protons.
p2, p1 and p0 represent 80.2--101 MeV, 50.8--67.3 MeV and  13.8--16.9 MeV protons for SOHO SEP events
and 60--100 MeV, 40--60 MeV, and 10--12 MeV protons for STEREO SEP events.
From the eighth to thirteenth columns are the onset times of type III, metric type II, decameter-hectometric (DH) type II, CME speed and source location and CME heights at the e1 release times. The fourteenth column denotes the observing spacecraft and the fifteenth is the connection angle of SEP to the spacecraft.

In this work, we used the TSA method to infer particle release times and 
8.33 minutes have been added to the release times in order to directly compare with electromagnetic emission onsets.
Here the inferred SEP release times indicate when the particles are injected onto the field line connecting to the observer.
To avoid the large background effect, we set the SPR time as null '---:---' when a onset level is greater than 10\%.
%For the SEP events with the onset level $>$ 1\% but less than or equal to 10\%, we applied equation (3) to correct the background effect by choosing a normalized onset level as $1\%$ of peak flux in all data set.

\subsection{Time Differences between Electron and Proton Release Times}

\begin{figure}
 \noindent\includegraphics[width=0.9\txw]{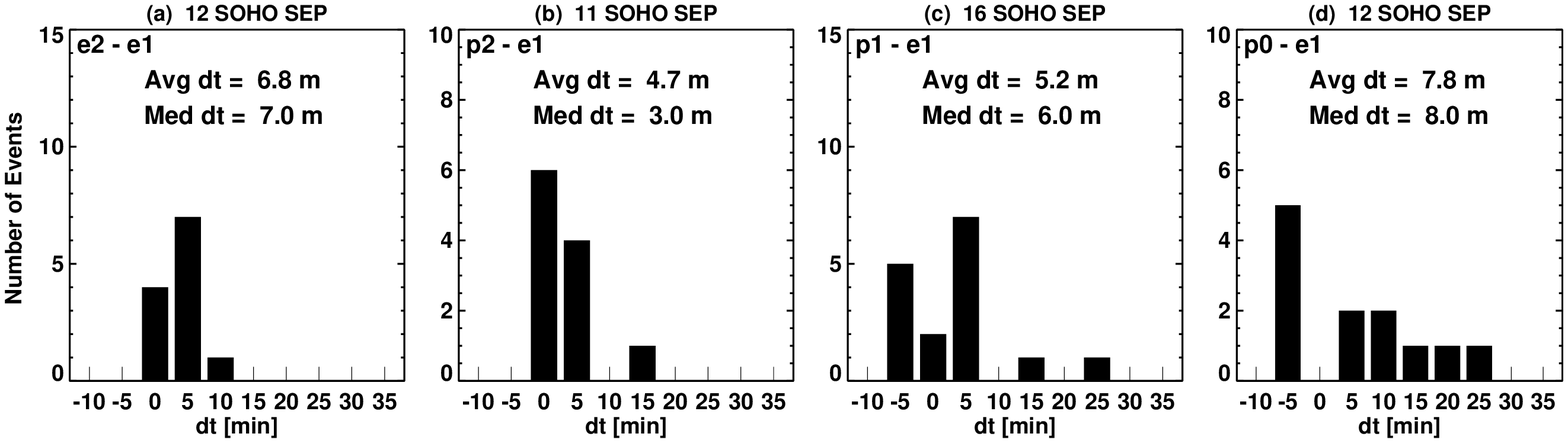}
% \noindent\includegraphics[width=0.7\txw]{Fig0_1b.eps}
  \caption{ Histograms of time differences for  17 SOHO SEP events between: (a) e2 and e1 SPRs, (b) p2 and e1 SPRs, (c) p1 and e1 SPRs and (d) p0 and e1 SPRs.}
\label{SOHO-SEPdt}
\end{figure}

Figure~\ref{SOHO-SEPdt} shows histograms of time differences, $dt$, for the 17 SOHO SEP events between: (a) e2 and e1 SPRs, (b) p2 and e1 SPRs, (c) p1 and e1 SPRs and (d) p0 and e1 SPRs, where  $dt = t_{SPR}(e2,p2,p1,p0) - t_{SPR}(e1)$, i.e., $dt$ is positive when the e2 and proton SPR times are delayed from the e1 SPR times. 
In panel (a), the e2 release times are found to be systematically larger than the e1 release times with an average of 6.8 min.
11 of the 12 events ($\sim$91\%) have $dt < $ 10 min and one event (event 6) has a delay of 14 min.
In panel (b), the p2 release times are delayed from the e1 release times with an average of 4.7 min. 
%in a range of 1 min to 8 min.
10 out of 11 events ($\sim$91\%) have $dt < $ 10 min and one event (event 6) has a delay of 19 min.
The p1--e1 SPRs in panel (c) show similar delays as the p2--e1 SPRs, ranging from -3 min to 25 min,
with an average of 5.2 min. 
For the p0 protons, 7 of 12 events ($\sim$58\%) have $dt < $ 10 min and five SEPs (event 3, 4, 6, 8 and 9) have large delays of  $\ge$ 10 min. 
Among these five SEPs, events 3, 6 and 8 are weak with small flux increases in e2 and p2.
Event 4 is associated with a high latitude source and events 4 and 9 shows a large proton scattering effect(see Sections 3.5).

\begin{figure}
 \noindent\includegraphics[width=0.9\txw]{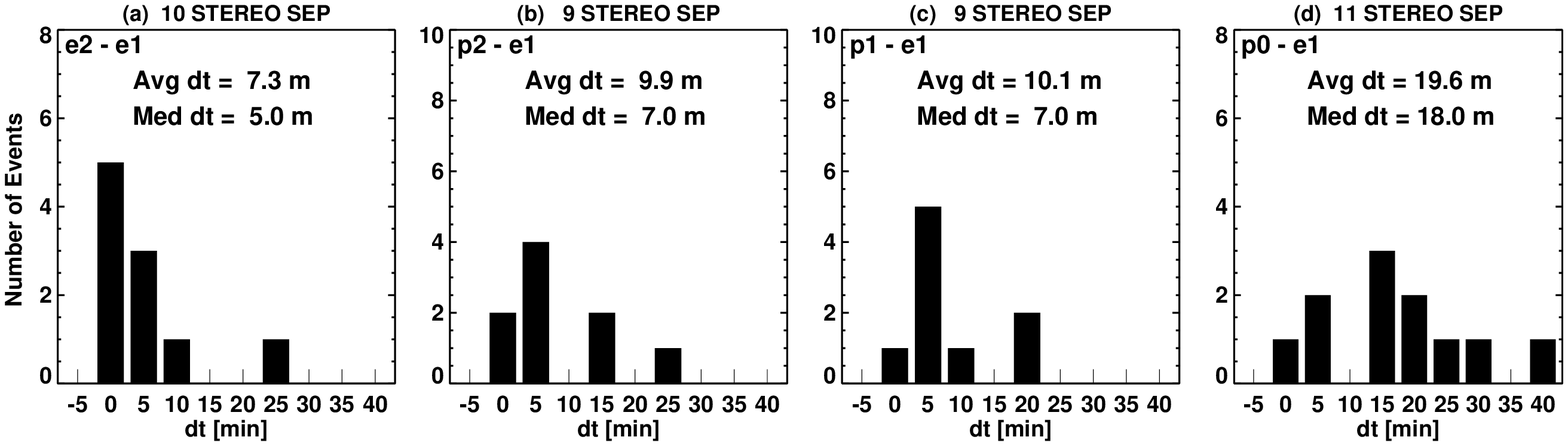}
  \caption{Histograms of time differences for the 11 STEREO SEP events between: 
(a) e2 and e1 SPRs, (b) p2 and e1 SPRs, (c) p1 and e1 SPRs and (d) p0 and e1 SPRs.}
\label{ST-SEPdt}
\end{figure}

Figure~\ref{ST-SEPdt} shows histograms of time differences, $dt$, for the 11 STEREO SEP events between: 
(a) e2 and e1 SPRs, (b) p2 and e1 SPRs, (c) p1 and e1 SPRs and (d) p0 and e1 SPRs.
Figure~\ref{ST-SEPdt} displays a similar trend as Figure~\ref{SOHO-SEPdt}.
%and the average $dt$ in Figure~\ref{ST-SEPdt} are slightly larger than in Figure 2.
Two events (event S5 and S10) in the e2--e1 SPRs and three events (event S5, S6 and S9) in the p2--e1 SPRs show a large delay of 12-28 min. 
%All these four events have suffered proton scattering effects (see Section 3.5),  
Among these three events, S6 is associated with a low CME speed with CA of 4$^\circ$ 
%(S6) and 61$^\circ$ (S8), 
and S5 and S10 have no associated metric type II but DH type II bursts, indicating a later shock formation time.
For the p0 protons, 8 events present a broad 10-41 minute delay relative to the e1 release times, and 3 events (events S1, S4 and S8) have $dt < $ 10 min. Among the 8 events with larger delays, 4 events (S2, S3, S9 and S11) have experienced strong scattering effects (S7 has no data availalbe for the VDA) and 3 events (S5, S6 and S10) have delayed proton release times.

\begin{figure}
 \noindent\includegraphics[width=0.9\txw]{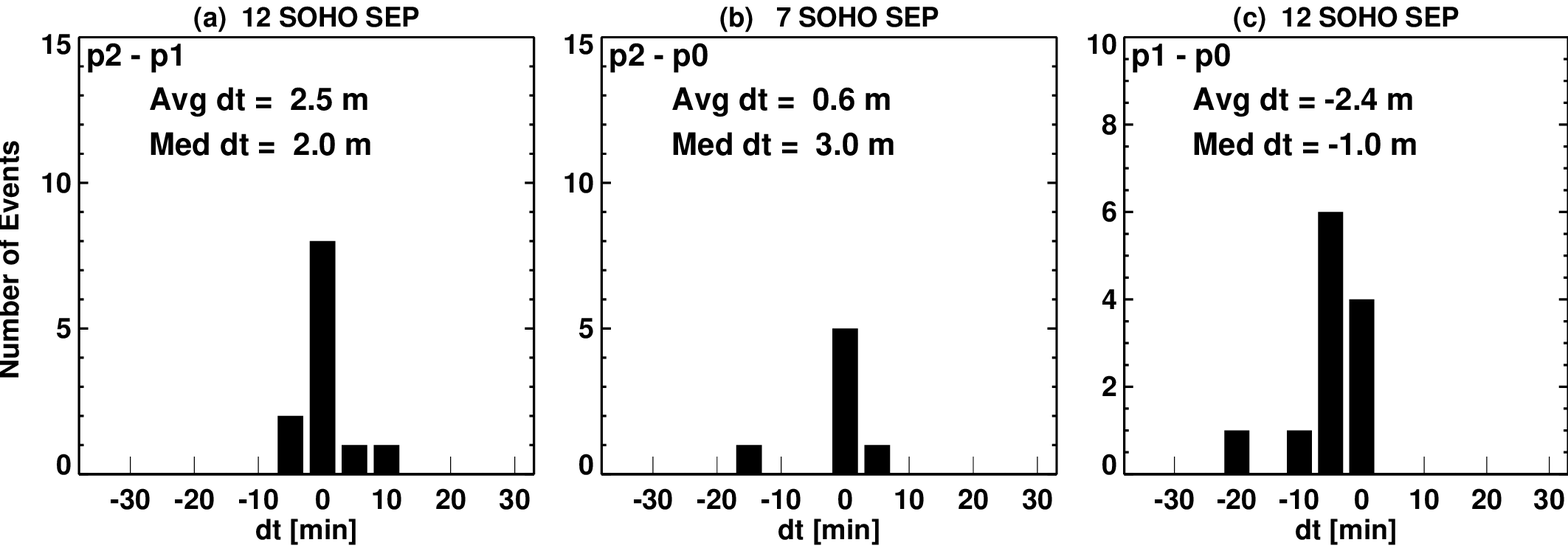}
% \noindent\includegraphics[width=0.7\txw]{Fig0_1b.eps}
  \caption{ Histograms of time differences for SOHO SEP events between: (a) p2 and p1 SPRs, (b) p2 and p0 SPRs, and (c) p1 and p0 SPRs.}
\label{SOHO-eng}
\end{figure}

\begin{figure}
 \noindent\includegraphics[width=0.9\txw]{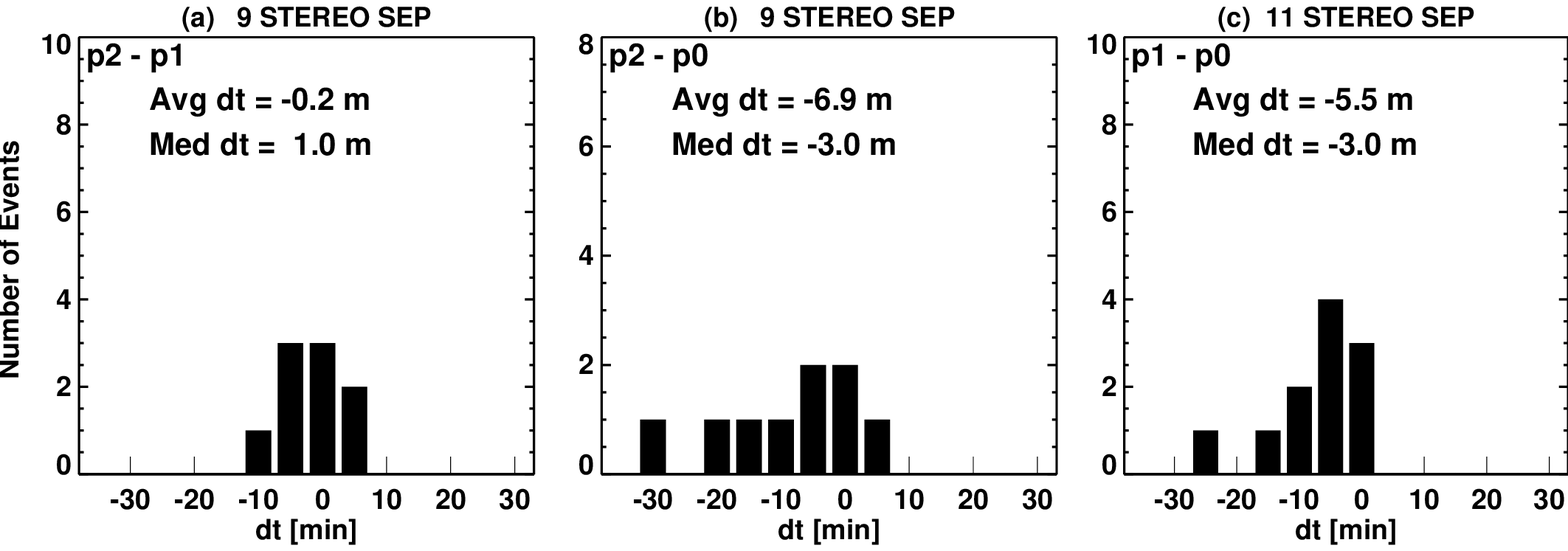}
  \caption{Histograms of time differences for STEREO SEP events between: 
(a) p2 and p1 SPRs, (b) p2 and p0 SPRs, and (c) p1 and p0 SPRs.}
\label{ST-eng}
\end{figure}

We plot histograms of time differences, $dt$, between: (a) p2 and p1 SPRs, (b) p2 and p0 SPRs and (c) p1 and p0 SPRs in Figure~\ref{SOHO-eng} (SOHO events) and Figure~\ref{ST-eng} (STEREO events).  
Both Figure~\ref{SOHO-eng} and Figure~\ref{ST-eng} show that the p2 protons have similar release times as the p1 protons with an average $dt$ of $\sim$  2.5 min and -0.2 min.
For the p0 protons, %half (9 of 18) of the SEP events have earlier p3 SPR times with average $|dt|$ of 9.9 min.
there are 7 SOHO SEPs and 5 STEREO SEPs with delays of the p2--p0 SPRs within 5 min, 
%which are small scattering events with electron anisotropy values A>0.6; 
two SOHO events 4 and 9 and four STEREO events S2, S3, S9 and S11 show proton scattering effects, where p0 appeared to be 
released later than p2 and p1, i.e., $dt$ = p2-p0 or p1-p0  is negative.

\subsection{Time Differences between Electron Release Times and Radio Emission Onset Times}

\begin{figure}
 \noindent\includegraphics[width=0.9\txw]{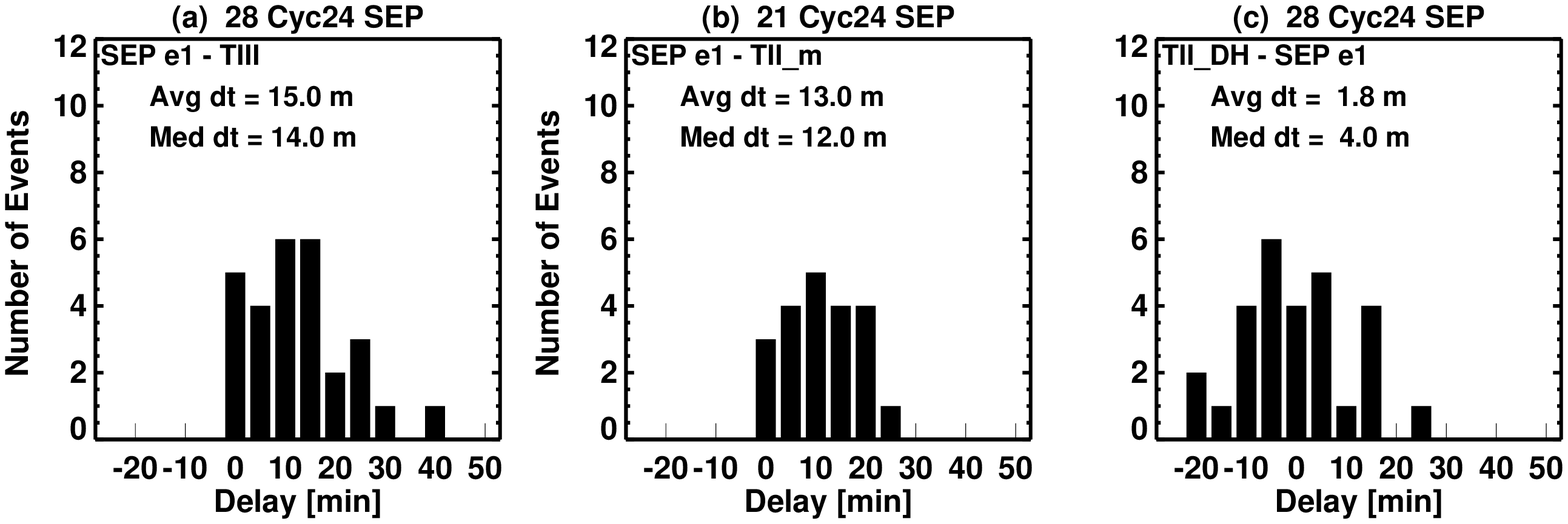}
  \caption{ Time differences between (a) e1 SPR times and type III onset times, (b) e1 SPR times and metric type II onset times,
and (c) e1 SPR times and DH type II onsets.}
\label{Tii_SEP}
\end{figure}

\begin{figure}
\noindent\includegraphics[width=0.7\txw]{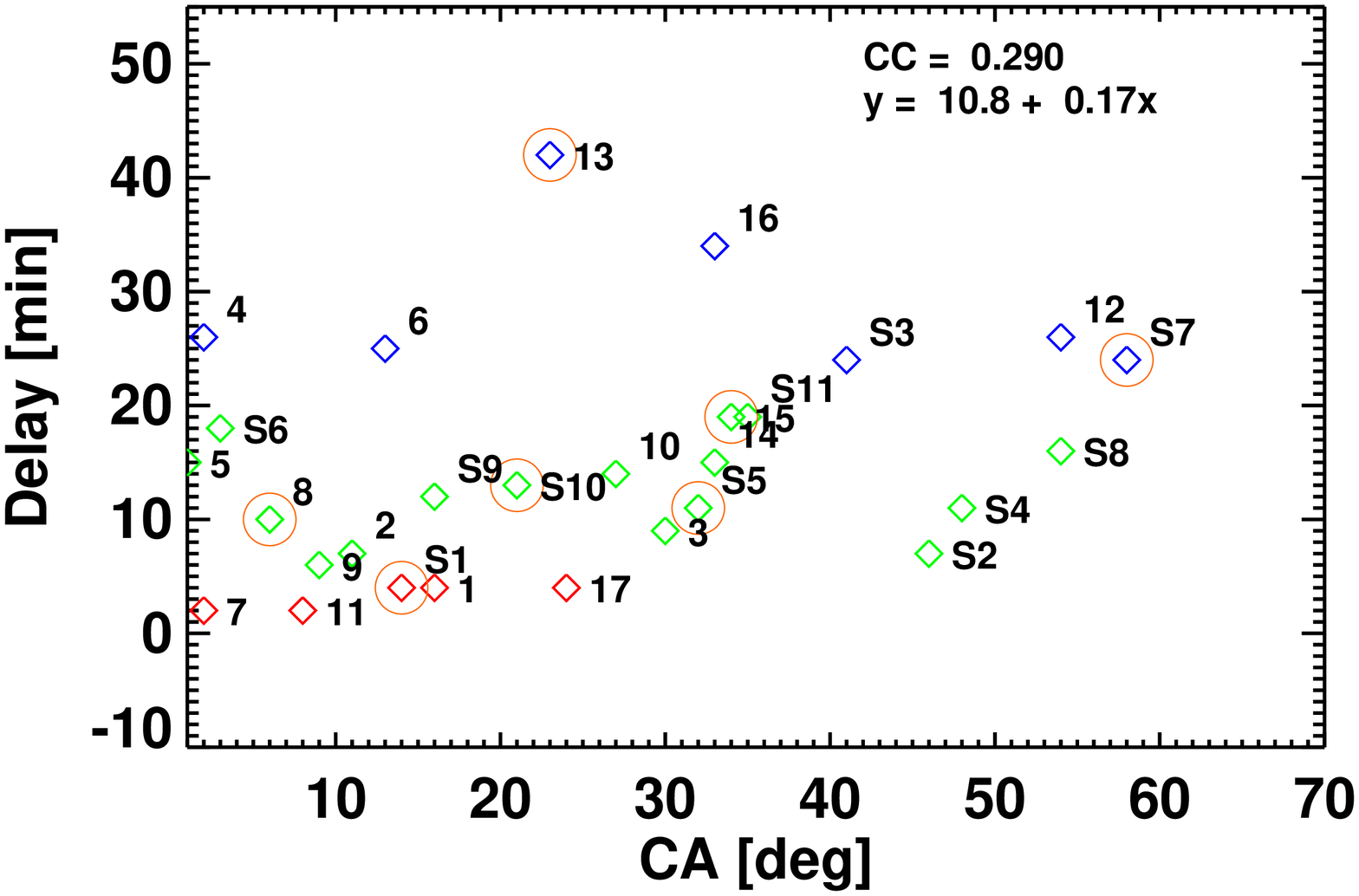}
  \caption{ Delays between the e1 release times and type III onsets as a function of CAs. Red, green and blue colors mark three groups with delays $dt \le 5$ min, $5 < dt < 20$ min, and $dt \ge 20$ min. Circles denote the events lack of associated type II bursts}
\label{CA_dt} 
\end{figure}

Figure~\ref{Tii_SEP} plots histograms of time delays between (a) e1 SPR times and type III onset times, (b) e1 SPR times and metric type II onset times,
and (c) e1 SPR times and DH type II onsets.
The e1 release times are found to be 2--42 min delayed from type III onset times, and similarly 3--25 min from metric type II onset times.
There are 5 events (1,7,11, 17 and S1) with delays $<$ 5 min %which released at CME heights $<$ 3 Rs;
and 7 events (4, 6, 12,13, 16, S3 and S7)) with delays $>$ 20 min.
Most of events (59\%, 16 of 27) have the delays ranging from 6--19 min.
Among the former 5 events (1,7,11,17 and S1), 4 of them are associated with metric type II burst except S1, which has a DH type II detected 5 min later than the e1 SPR time.
There are in total 7 events (see Figure~\ref{CA_dt}) having no associated metric type II bursts but all of the 28 SEP events are associated with DH type II bursts. 

Figure~\ref{CA_dt} presents correlation of CAs and time differences, DT, between the e1 SPR times and metric type III onsets. Red, green and blue colors mark three groups 
with delays $dt \le 5$ min, $5 < dt < 20$ min, and $dt \ge 20$ min.
%21 events out of 28 SEP events are used to perform the correlation analysis since there are 7 SEP events which have no associated metric type II bursts. %and 1 SEP with no e1 SPR time.
From Figure~\ref{CA_dt}, we can see that there is a poor correlation between CAs and DTs with correlation coefficient CC = 0.167 and the data points are widely spread in the whole plot.
Event S8 has the second largest CA of $ -54^\circ$ but a relatively small delay of 16 min.
This SEP event was associated with a M6.5 flare at N09E12. STB/HET, SOHO/EPHIN and ERNE, and GOES have all detected a rapid rise of SEP fluxes, despite of relatively large CA to STB  ($61^\circ$) and  to SOHO ($70^\circ$).
This is one of the longitudinally wide-spread SEP events (Richardson et al, 2014) and will be studied in our future work.
On the other side, event 4 is the SEP event which has a small CA of  $2^\circ$ but a large delay, associated with a M3.7 flare at active region (AR) 11164 at N31W53.
The type III onset, metric type II onset and the 0.25--0.7 MeV electron release time are 19:52 UT, 19:54 UT and 20:21 UT respectively.
Although this is a well-connected SEP event to SOHO, there is a large delay of 24 min between the e1 release and metric type II onset. 
A likely reason is that although this SEP event is well-connected to SOHO in longitude, however, its large source latitude (N31) and relatively small CME width kept it poorly-connected to the ecliptic plane \citep[cf][]{Gopalswamy2014}.

%, (c) CME speeds, and CME heights at the e1 SPR times.
%9 of 21 ($\sim 43\%$) events in panel (a) have $|dt| \le $ 10 min with an average delay of 7 min.
%12 of 21 ($\sim 57\%$) events have a broad 10-34 min delay with an average value of 19 min.
%The average delay for the total 21 events is $\sim$ 14 min.

%although we can still not distiguish if these mechanisms are flare-related or shock-related or both.

\subsection{SEP Electron Anisotropy}
The TSA method assumes that the SEP particles have propagated scatter-free at zero pitch angle along the magnetic field line, large errors may present in the TSA method for events with strong scattering. 
To estimate the scattering effect of the first arriving particles, we compute the electron anisotropy using Wind/3DP data for the SOHO events (we used ACE/EPAM electron data for events 1 and 2 when there was a data gap in Wind/3DP) and SEPT data for the STEREO events. 
The solid state telescope (SST)  Foil pitch angle distributions (SFPD) for Wind 3DP electrons (available at 
\url{ftp://cdaweb.gsfc.nasa.gov/pub/data/wind/3dp/3dp_sfpd/}) returns a velocity distributions function containing 7 energy bins from $\sim$ 27 keV to 520 keV  and 8 pitch angle bins roughly  covering pitch angles from 0-180$^\circ$. Note that the covered pitch-angles can vary from distribution-to-distribution since the automated CDF routine tends to remove all the direct sun/anti-sun directions to avoid X-ray and EUV contamination (see  \url{http://cdaweb.gsfc.nasa.gov/misc/NotesW.html#WI_SFPD_3DP}).
The SEPT instrument provides 45--400 keV electron measurements. It consists of four identical telescopes which cover four viewing directions: SUN (along the nominal Parker spiral to the Sun), ANTI-SUN (away from the Sun), NORTH and SOUTH.
In this Section, the anisotropy of the protons are not computed due to the lack of anisotropy data in ERNE and HET. Instead, we use the VDA to evaluate their scattering effects in Section 3.5.

The anisotropy of a SEP event is defined as

\begin{equation}
A=\frac{3\int_{-1}^{+1} I(\mu) \cdot \mu \cdot d\mu}{\int_{-1}^{+1} I(\mu) \cdot  d\mu} 
\end{equation}

where $I(\mu)$ is the intensity at a given pitch-angle direction and $\mu$ is the pitch angle consine.
Omnidirectional intensities were calculated by integrating second-order polynomial fits to 
the pitch-angle distribution of intensities using 1-minute averages (12-second for Wind/3DP) of the data.
To stabilize the fit during periods of poor pitch-angle coverage, an artificial point was added to the pitch-angle distribution to 
fill the uncovered range \citep[cf][]{Droge14}.

\begin{figure}
\noindent\includegraphics[width=0.7\txw]{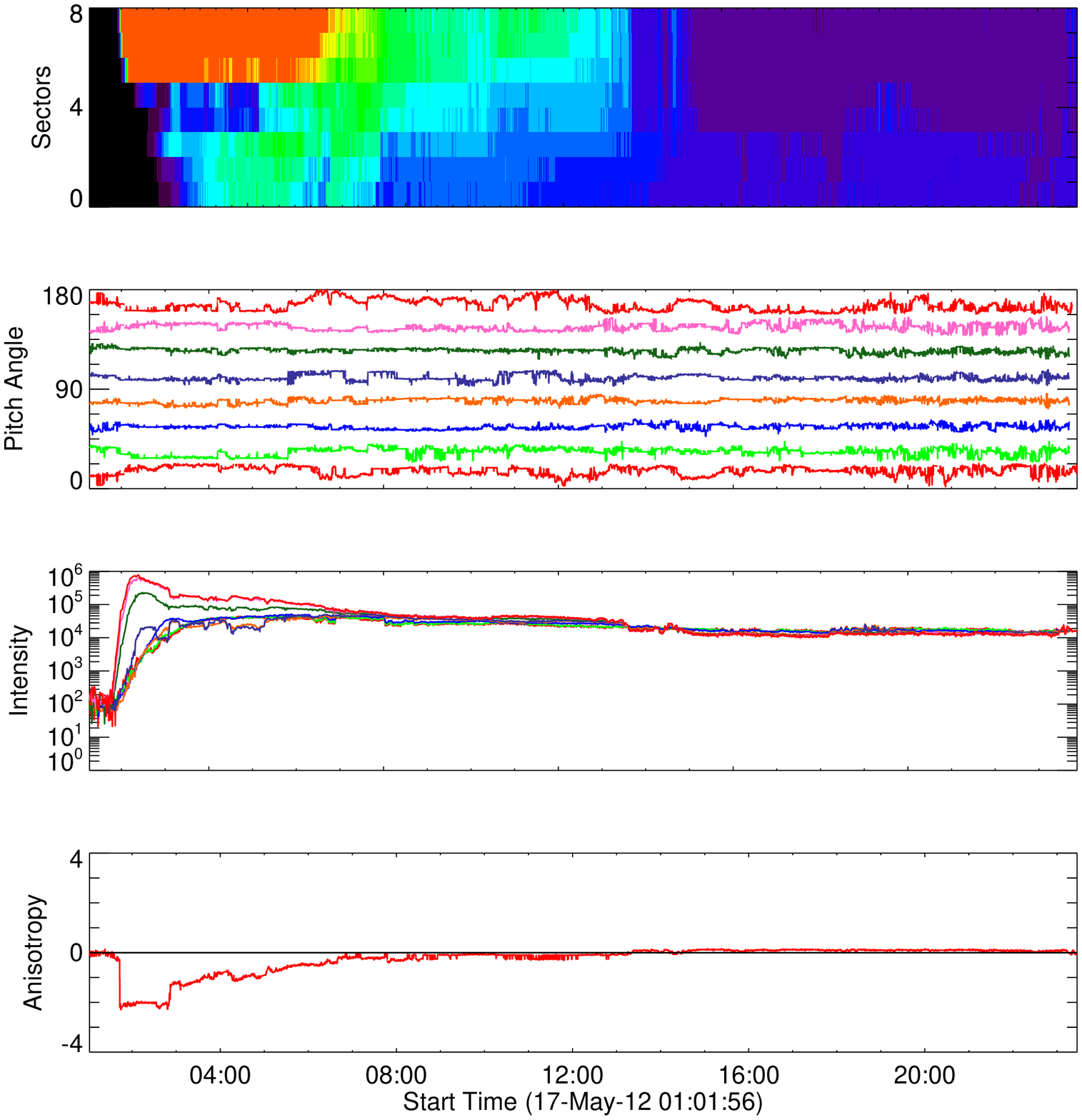}
  \caption{Anisotropy and intensity time profiles of the SEP event on May 17 2012 observed by Wind/3DP. }
\label{aniso} 
\end{figure}

Figure~\ref{aniso} shows Wind 3DP SFPD measurements on the May 17 2012 SEP event, which serves as a good example
of a strong anisotropic event. The upper panel shows the time series of the intensity in color coding as a function of pitch
angle bins. The middle panel shows the 65 keV electron 8-bin intensity measured by the SST telescope. 
The third panel shows the anisotropy as computed from the pitch angle distribution measurements.
The anisotropy reaches a maximum of 2.23 at 01:44 UT during the onset of this event.

In Table 2 column 2 we list the maximum anisotropy for the 27 SEP events (data are not available yet for the STEREO SEP event on 2014 February 25).    
The obtained anisotropies range from 0.27 to 2.97 (absolute values), which are similar to those computed from ACE/EPAM data in \citet{Dresing14}.
6 out of 27 SEPs have relatively strong anisotropies with A $>$ 2.0, 16 events have 1.0 $\ge$ A $\le$ 2.0,  
and 5 events have relatively weak anisotropies with A $<$ 1.0.
The obtained anisotropies suggest that most of electrons with finite pitch angles still experienced certain degree of scattering although the first-arriving electrons with $\mu \sim 0$ are generally propagated with less scattering.
The uncertainties of e2 and e1 brought by the scattering effects are 6 min and 8 min, respectively, for a path length of 2 AU.

%one event 16 has a weak anisotropy with A $<$ 0.65, and two events (9, 14) have almost no anisotropy with A $<$ 0.6.
%In general, the first-arriving electrons with $\mu \sim 0$ have been propagated relatively scatter-free and a good magnetic connection to the SEP source region have been present.
%are consistent with the focused transport model. 
%For $\sim$ 93\% (25 of 27) SEP events under study with CAs $\le$ 58$^\circ$, the effect of focusing of near-relativistic electrons in the interplanetary magnetic field has overcome the degree of scattering at magnetic fluctuations, and propagated nearly scatter-free.

\subsection{The Proton VDA Release Time}

In this section, we carry out the VDA to estimate the proton scattering effects and compare the proton VDA release times with electron release times.
The VDA analysis was based on 1 min time resolution ERNE and HET (LET) proton data with energy channels between 10 MeV and 100 MeV. 
The VDA onset times are determined based on the fix onset level (see Figure 1) for all energy channels, which is selected to be the minimum background level of all analysis channels. 
To avoid high-background effects or errors brought by background variation, 
we have excluded the channels with background levels $>$ 10\% and channels with a slowly rising background (see details in Figure 9 and Figure 10).
In addition, to avoid energy-dependent scattering effect, we carried out the VDA by using either high $\sim$ 
50--100 MeV or low $\sim$ 10--50 MeV energy channels only, depending on data availability.
The energy range used in the analysis has been listed in column 8 of Table 2 .

   \begin{figure}[t!]
       \mbox{
    \includegraphics[width = 0.98\txw,height = 0.90\txw ]{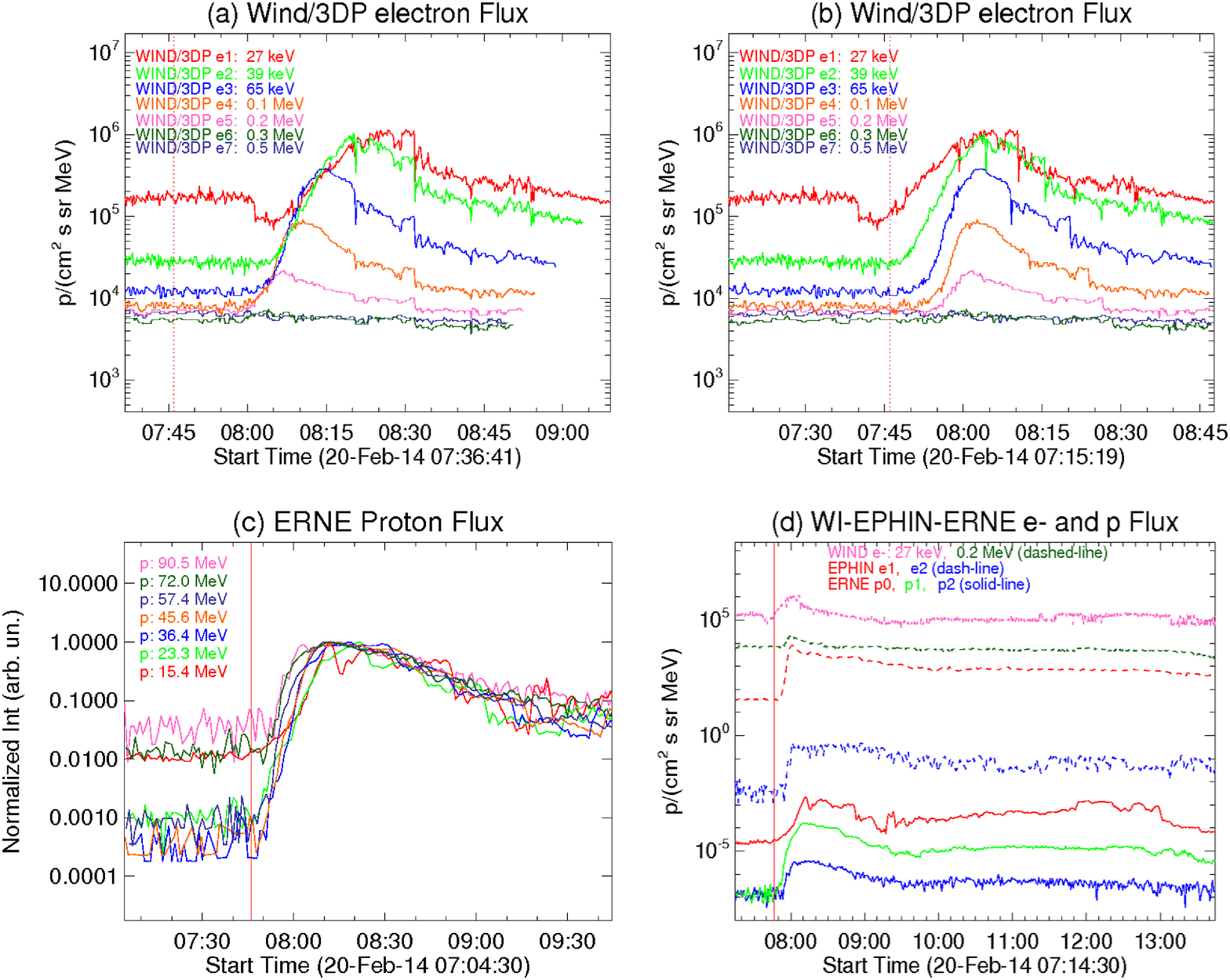}
    }\par

  \caption{SEP intensity on 2014 February 20: (a) 0.027 -- 0.5 MeV electron intensity from Wind/3DP,  (b) time-shifted 0.027 -- 0.5 MeV electron intensity from Wind/3DP,  c) 15.4 -- 90.5 MeV normalized and time-shifted intensity from ERNE and d) over-plotted electron and proton intensity from Wind/3DP, EPHIN, and ERNE.}
\label{Fig0220}
 \end{figure}

%    }\par         \mbox{
%   \includegraphics[width=0.32\txw, height = 0.32\txw ]{Fig0220_c.eps}
%    \includegraphics[width=0.32\txw, height = 0.32\txw ]{Fig0220_d.eps}

%    \includegraphics[width=0.35\txw, height = 0.35\txw ]{Fig0220_d.eps} where the SEP intensity has been normalized to the peak flux and the SEP time has been shifted with the travel time of SEP with 1.25 AU path length

Figure~\ref{Fig0220} shows an example of SEP event on February 20 2014.
The 2014 February 20 SEP is a strong anisotropic event with a maximum anisotropy A = 1.94. 
This SEP event was associated with a M3.0 X-ray flare at S15W73 and a halo CME with speed of 948 km/s. The observed metric and DH Type II, and type III onsets are 07:45 UT and 08:06 UT, and 07:46 UT, respectively.
The event CA is $-24^{\circ}$ and GOES observed a small SEP intensity of 22 pfu. 

Figure~\ref{Fig0220} (a) plots the 12-second electron intensity in 27--520 keV energy channels from Wind/3DP SFPD from the solar direction. 
A clear velocity dispersion in the peak flux is visible. %has been seen in Figure~\ref{Fig0220} (a).
The velocity dispersion at the onset shows an instrumental effect:
the intensity at lower energy channels were contaminated by higher-energy channels. This occurred when the high-energy electron lost only a fraction of its energy in the detector, a count at a lower energy was recorded resulting in too early onset times.
The early onset effect at low energy can be more clearly seen in Figure~\ref{Fig0220} (b), where the electron time profiles have been shifted by the travel time of SEPs with 1.25 AU path length.
%Due to soft spectrum intensity, not many electrons were seen above 300 keV, thus the highest channel 180 keV channels gives the true onset time.

\begin{figure}[t!]
       \mbox{
    \includegraphics[width=0.7\txw, height = 0.7\txw ]{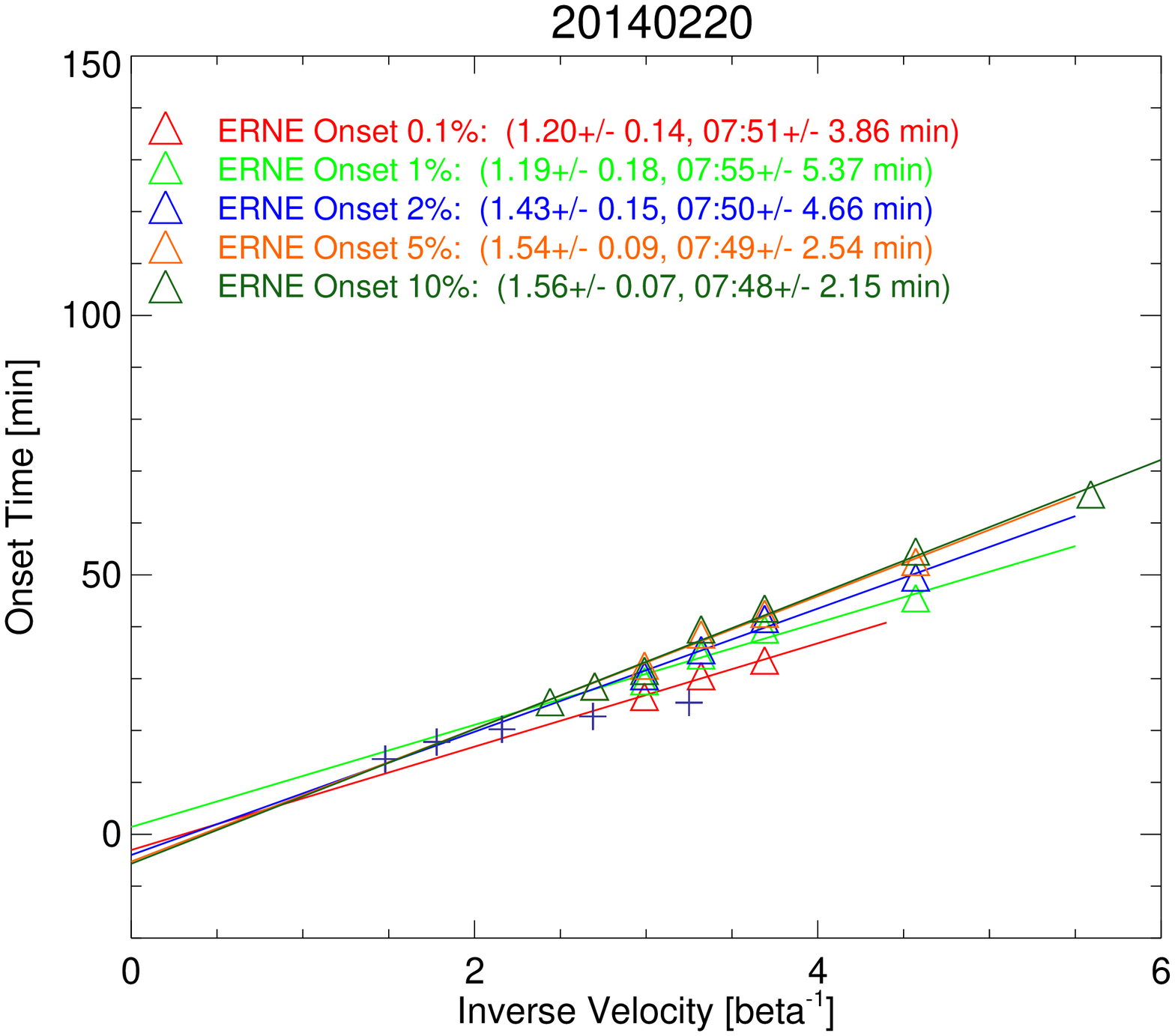}
    }\par
  \caption{The VDA based on onset times at 0.1\%, 1\%, 2\%, 5\%  and 10\% of the peak value.}
\label{Fig0220a}
 \end{figure}

Figure~\ref{Fig0220} (c) plots the 1-minute proton intensity from ERNE, and (d) superimposed intensity profiles from Wind/3DP electrons, EPHIN electrons and ERNE protons on 2014 February 20.
For easy comparison, in Figure~\ref{Fig0220} (c) the intensity profiles have been normalized to the peak values and the travel times have been subtracted with 1.25 AU path length. The red vertical solid line in the figure indicates the type III burst onset time. 
%Note that a similar false early onset effect was shown in the 15.4 MeV proton channel. We have excluded this channel 
Note that a slow rise of the background flux was shown in the 15.4 MeV proton channel before the sharp rising phase,
to avoid the onset uncertainty, we have excluded this channel in the analysis based on onset levels less than 10\% of the peak value (see Figure~\ref{Fig0220a}).

Figure~\ref{Fig0220a} presents the VDA results based on onset times at 0.1\%, 1\%, 2\%, 5\%  and 10\% of the peak value.
%The VDA gave the first arriving (at 0.1\% level) proton release time at 07:51 UT  with an uncertainty of $\pm$ 3.9 min, and 
%For later arriving protons at 10\% of the peak level, the release time is at 07:50 UT  with an uncertainy of $\pm$ $\sim$ 2.3 min, and a path lenght of 1.53 AU with an uncertainy of $\pm$ 0.1 AU.
The results show that the first arriving protons at 0.1\% onset level propagated nearly scatter-free with a path length of 1.2 $\pm$ 0.14 AU. 
The later arriving protons  at onset level $>$ 5\% present a larger scattered path length with a path length of $\sim$ 1.5 AU.
However, although the scattered path lengths increase as the onset levels increase, the VDA proton release times remain roughly the same within 7 min of uncertainty.
The TSA release time for 180 keV electrons from Wind/3DP is at 07:52 UT and for the e1 and e2 electrons from EPHIN are at 07:50 UT and 07.54 UT, thus
no significant differences between proton and electron SPR times are found for this SEP event.

   \begin{figure}[t!]
       \mbox{
    \includegraphics[width=0.7\txw, height = 0.7\txw ]{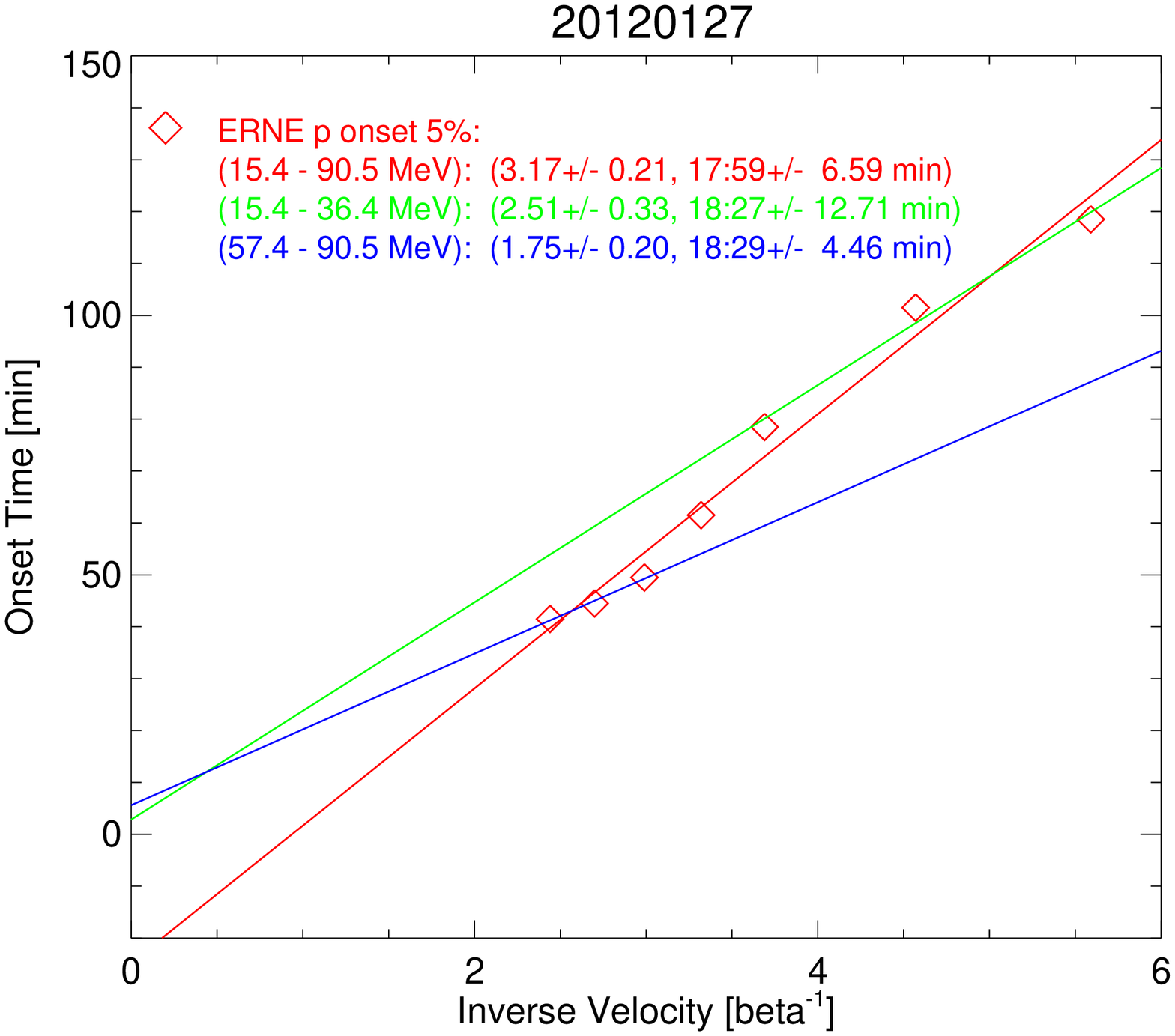}
    }\par
  \caption{The VDA based on 57.4--90.5 MeV (blue), 15.4--57.4 MeV (green) and 15.4--90.5 MeV energy channels.}
\label{Fig0127a}
 \end{figure}

Caution has to be taken with the VDA due to high background effect and energy-dependent scattering effect.
The energy-dependent scattering effect becomes more important for cases when there is a large amount of scattering. 
For such cases, the VDA using the energy range from 10 to 100 MeV may yield a release time that is earlier than expected \citep{Diaz11}.
Figure~\ref{Fig0127a} shows a good example of such a SEP event on 2012 January 27 which has a weakest anisotropy with A = 0.27. 
%The SEP event was associated with a X1.7 X-ray flare at N27W71 and a halo CME with speed of 2508 km/s. The observed metric and DH Type II, and type III onsets are 18:04 UT and 18:17 UT, and 18:27 UT, respectively.
%It is a well-connected SEP event with CA of -9.
%but preceded by a large SEP event on 2012 January 23. This large SEP event has caused the intensity saturation in ERNE (excluded in the study) and is likely the reason causing a large amount of scattering in the following 2012 January 27 SEP.
As shown in Figure~\ref{Fig0127a}, the velocity dispersion onset times do not lie on a straight line but curved from high to low energy, showing an increasing scattered path length. 
The VDA based on energy 15.4--36.4 MeV (green) yielded a path length of 2.51 AU, which is $\sim$ 0.76 AU larger than that in 57.4--90.5 MeV (blue);
and the VDA using energy 15.4--90.5 MeV gave a too large path length of 3.17 AU and a unreasonable release time that is earlier than the type III onset.

%Our derived results are consistent with \citet{Diaz11}'s study, where they found in the worst scenario with mean free paths $\lambda_0$ = 0.10 AU, the VDA yielded a SPR error of $\sim$ 28 min earlier and a path length of $\sim$ 3.0 AU, $\sim$ 1.4 AU--0.6 AU larger than real values of 1.6 - 2.4 AU  for  100 - 10  MeV protons.

%In Talbe 2, we list the electron anisotropy, e1 SPR time, e2 SPR time and the VDA results 
%for the 28 SEP events. 
%Columns 3--4 are the TSA release times for the e1 and e2 electrons.
In Table 2, we list the VDA results along with the electron anisotropy, e1 SPR time, e2 SPR time for the 28 SEP events.
Entries with '---' (5 of 28) are cases where the path length values were outside the range of 1--3 AU due to high background level (HBG), ion contamination (IC), or cases when data are not available (NA). 
The obtained proton path lengths range from 1.18 to 2.51 AU. 
In general, the derived proton path length tend to be larger for weak anisotropic events than strong anisotropic events.
However, there are cases with strong electron anisotropies show large apparent proton path lengths due to the relatively high background levels (especially for the STEREO HET data).
%where the HET proton data show higher background levels than ERNE), as shown in Figure~\ref{Fig0220a}.
%it is aslo affected by the background level, as shown in Figure~\ref{Fig0220a}, higher onset levels tend to give larger scattered length.
15 out of 23 events have the path lengths $l \le$ 1.5 AU and 21 of 23 events have $l \le$ 1.65 AU except events 9 and S3, where event 9 (the January 27 2012 SEP event) has the weakest anisotropy A = 0.27.
By comparing the TSA release time with the derived release times from the VDA,  we obtain the maximum errors for the p2, p1, p0 protons of $\sim$ 6 min, 10 min and 32 min, respectively.
The p2 TSA release times have the smallest error as expected. The proton release times from VDA, $t_{SPR}(p_{vda})$, are found to be delayed from
the e1 SPRs by -1--30 min, and from  the e2 SPRs by -9--18 min.
$\sim$ 70\% (16 of 23) events have  the $dt_1 = t_{SPR}(p_{vda}) - t_{SPR}(e1)$  $<$ 8 min and seven events (3, 6,8,S5,S6,S9 and S10) have $dt_1 \ge $ 8 min.
13 (out of 19) events have $dt_2 = t_{SPR}(p_{vda}) - t_{SPR}(e2)$ within 6 min and 3 events (S6, S9 and S10) have $dt_2 >$  9 min.
In addition, there are 3 events (1, 7 and S2) having a negative $dt_2$, where events 1 and 7 suffered the X-ray contamination resuting in a large undertainty of $\sim$ 10 min in the e2 SPR times.

\section{Summary and Discussion}
\subsection{Summary}
By choosing the smallest CA among the three spacecraft, we derive and compare the high energy electron and proton SPR times using SOHO/EPHIN electron fluxes in  the 0.25--10.4 MeV channels, SOHO/ERNE proton fluxes in the 13.8--101 MeV channels, or in the similar energy channels of the SEPT and HET (LET) detectors on STEREO. Our main results are listed below.

\begin{itemize}

\item 
The e2 release times are found to be systematically larger than the e1 release times by an average of 6.8 min and 7.3 min, for the 12 SOHO SEPs and 10 STEREO SEPs, respectively.
Among these 22 events, three events (6, S5, and S10) have a large 10--28 min delay.

%\item 
%The p2 release times are delayed from the e1 release times by an average of 4.7 min and 8.2 min, for the 11 SOHO SEPs and 9 STEREO SEPs, respectively.
%There are three events (6, S5,and S6) showing a large 12--25 min delay. 

\item 
The p2 protons are shown to have similar SPR times with the p1 protons. The average delay between the p2--p1 SPRs are $\sim$ 2.5 min and -0.2 min, for the 12 SOHO SEPs and 9 STEREO SEPs, respectively. 
For the p0 protons, there are 12 SEP events showing small delays between the p2--p0 SPRs within 5 min and five events (9,S2, S3, S9 and S11) showing a large 10--32 min delay due to proton scattering effects.

%\item
%From the VDA, the protons and the e2 electrons in 17 of 19 SEP events appear to be released simultaneously within 9 min;
%and in 17 of 23 SEP events they appear to be released simultaneously with the e1 electrons within 5 min.
%There are 6 events (6, 8, S5, S6, S9 and S10) having a large 13--30 min delay between the VDA proton release times and the e1 (and e2) SPR times.
%Among these 6 events, 3 events (6, S5, and S10) also a large e2-e1 SPR delay.

\item
The proton VDA results show that protons are released simultaneously with the e1 electrons within 8 min for $\sim$ 70\%(16 of 23) SEP events, and the e2 electrons with 6 min for 13 of 19 events. 
There are $\sim$ 30\%(7 of 23) SEP events showing a delayed proton release time by $\sim$ 8--31 min.
Among these 6 events, 3 events (6, S5, and S10) also have a large e2-e1 SPR delay.

\item
%Most SEP (23 of 27) events under study have a strong electron anisotropy with A ranging from 0.9 to 2.97.
$\sim$ 65\% (15 of 23) protons events show a small scattered path length ($<$ 1.5 AU);
8 of 23 proton events have a large apparent path length  ($>$ 1.5 AU), part of reason is due to higher background levels in the STEREO HET data.

\item
The delays between e1 SPRs and type III onsets range from 2 min to 42 min. 
The CME heights at the e1 release times range from 2.1 to 9.1 Rs. 
From the CME heights, it is likely that the e1 electrons are accelerated by the CME-driven and/or flare shock waves rather than flare reconnections.

\end{itemize}

%The above results indicate that electron and proton acceleration are closely related and released at about the same time.
\subsection{Discussion}
\subsubsection{Association between Electrons and Protons}
Our results are consistend with \citet{Haggerty09}'s study, where they suggested that near-relatic electrons and the energetic protons are accelerated and released by essentially the same mechansim(s).
%although we can still not distiguish if these mechanisms are flare-related or shock-related or both.
\citet{Haggerty09} studied the injection times of near-relativistic electrons and non-relativistic protons for 19 electron beam events using ERNE 50-100 MeV proton and EPAM 38-315 keV electron data, and found that
 11 of the 19 events (60\%) are statistically consistent with zero delay between
the proton and electron injection within the uncertainty of $\sim$ 3min. 
The remaining 8 events show a broad 5-25 minute delays of the protons relactive to the electron injections.
%\citet{Haggerty09} studied the electron assocation with 2-5 MeV proton.
They also compared the peak intensity of 175--315 keV elections with that of 1.8--4.7 MeV protons from ACE/EPAM and found a good correlation in the peak intensity of electrons and  protons. 
%They restricted their study to beams and the prompt component (whereas numerous other studies mix together non-beam and the ESP part of the event) and argued the correlation suggested that the near-relatic electrons and the energetic protons are accelerated and released by essentially the same mechansim(s).

\begin{figure}
\noindent\includegraphics[width=0.7\txw]{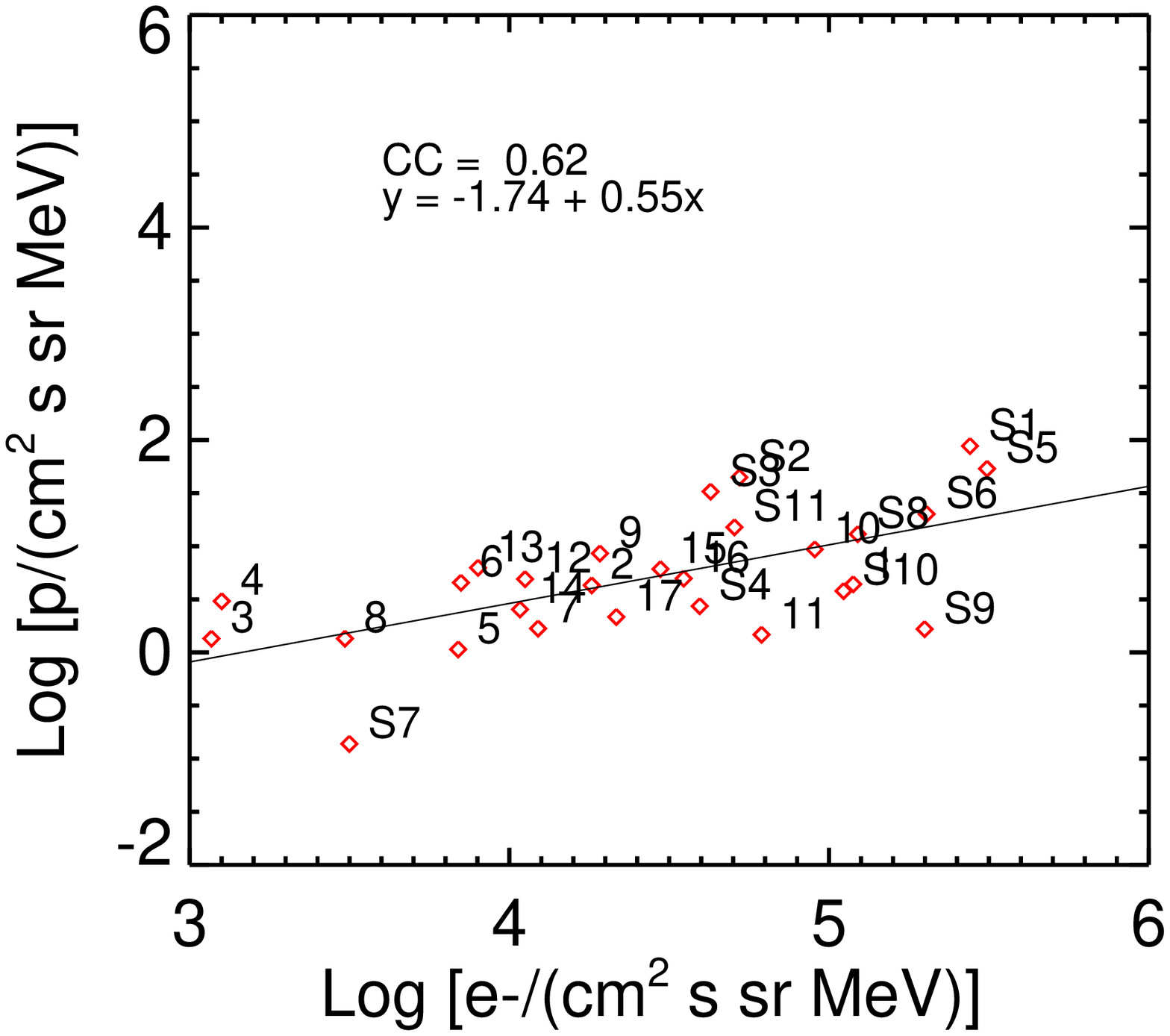}
  \caption{ Logarithmic peak intensity correlation between the e1 electrons and the p0 protons.}
\label{int_corr} 
\end{figure}

   \begin{figure}[t!]
       \mbox{
    \includegraphics[width=0.98\txw, height = 0.45\txw ]{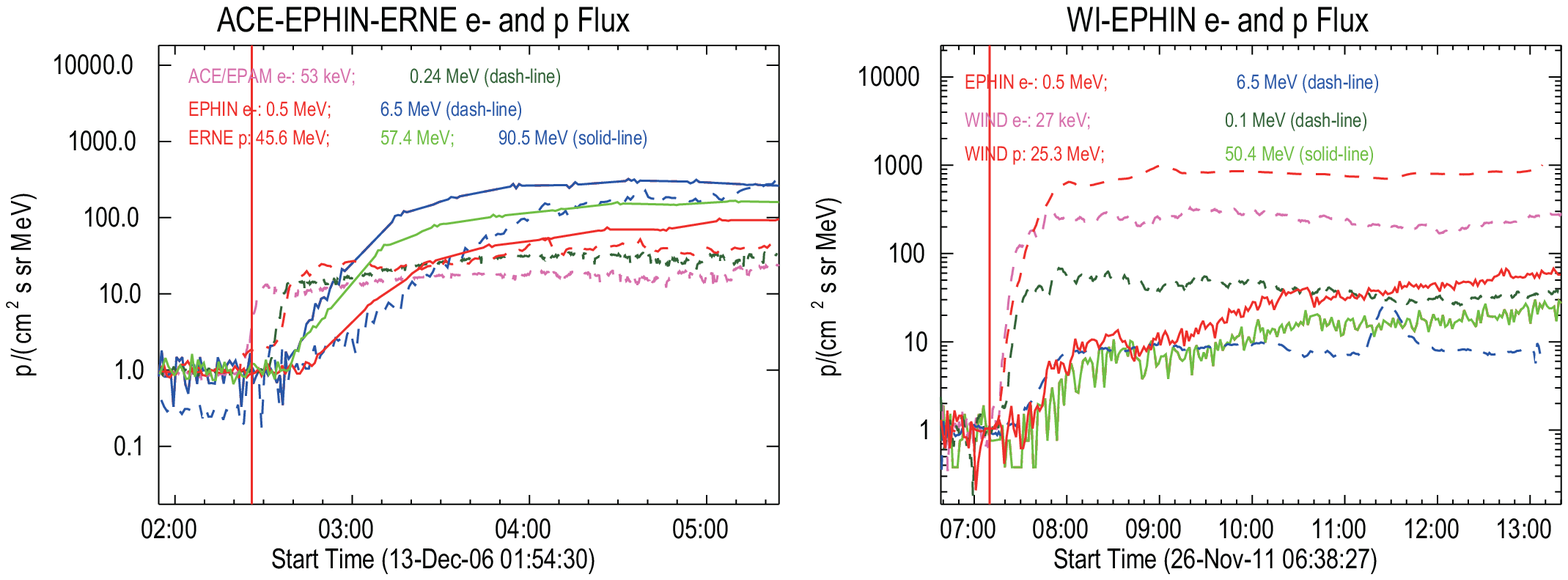}
    }\par
  \caption{ (Left) over-plotted electron and proton intensity from ACE/EPAM, EPHIN, and ERNE on December 13 2006.
  (Right) over-plotted electron and proton intensity from WIND/3DP electrons, EPHIN, and  WIND/EPACT protons on November 26 2011. The intensity has been normalized to the background flux level
  for easy comparison.}
\label{figall}
 \end{figure}

Among the 28 SEP events under study, we found similar correlations between the peak intensity of e1 electrons and p0 protons, as shown in Figure~\ref{int_corr}. 
Futhermore, the profiles between different spices are found to be very similar to each other although not identical, as shown in Figure~\ref{figall}.

Our results support the conclusion that near-relativistic electron and  high-energetic proton acceleration are closely related to each other.
On the other hand,  how the intensity profiles evolve with time, which result from the transport-modulated SEP particle accelerations at an evolving CME-driven shock, is not well understood.
For example, at the SEP rise phase, it is not well understood why the e2 electrons are the last to reach their peak value for event 1 (left in Figure~\ref{figall}); 
while in the second example (event 8, right panel in Figure~\ref{figall}), the e2 electrons reach the plateau before the protons.

\subsubsection{Direct Shock Accleration vs Tranverse Transport}
Besides simultaneously released electron and proton events, there are seven events showing large delays of 8--31 min between proton release times  $t_{SPR}(p_{vda})$ from VDA and e1 SPRs.
These events are SEPs with small e2 and p2 intensities. 
%which also show a large delay between the e1 SPR time and DH type III onset times. 
%They were released at times close to or later than the DH type II onset times.
Three possible reasons may account for these large delays:
1) the late formed shocks at high altitudes around DH type II onset times;
2) longer times needed  for the evolving shocks to be intense enough to produce high-energy SEPs after DH type II onsets;
3) times  needed for shocks in SEP events with large CAs to reach the magnetic connection footpoint to the observer.
Among the above 7 events, events 8 and S6 have small CAs of 6$^\circ$ and 3$^\circ$, events 3 and S5 have large CAs of 30$^\circ$ and 32$^\circ$, and
the other 3 events (3, 6, and S10) have intermediate CAs of 13-21 $^\circ$.
6 of these events have the similar e1 SPRs with the DH type II onsets within 5 min ( and a large 13-26 min delay between $t_{SPR}(p_{vda})$ and metric type II onsets) except event S10.
The obtained timing comparising results are consistent with one (or two) of the above three hypotheses. 
Rouillard et al. (2012) investigated the 2011 March 21 SEP event using STEREO and SOHO observations. By tracking the CME shock lateral expansion they demonstrated that the delayed solar particle release times are consistent with the time required for the shock to propagate to the magnetic footpoint connecting to the observer.
On the other hand, for large CA and/or high latitude SEPs, an alternative (or contributing) explanation is that the delay between the SEP release and electromagnetic emissions is caused by the propagation times needed for the SEP particles to transport across the field line to the connection footpoint of the observer \citep[e.g.][]{Dresing12, Qin13, Laiti15}.
% and M. Zhang, private communication, 2015).
It is possible that both direct shock acceleration and cross-field propagation of SEPs play roles in the formation of SEP intensity time profile.
At an evolving CME-driven shock near the Sun, many factors such as the shock obliquity, the compression ratio and transport parameters may affect the SEP intensity,  further investigations are needed.

\subsection{Conclusion}
Our results suggest that near-relativistic electron and high-energy proton acceleration are closely related to each other. There exists a good association between high-energy electron and proton release time, intensity peak values and time profiles. %CME heights at the e1 SPR times favor the shock accelerations over flare reconnection accelerations for the SEP events under study. 
For small intensity SEP events, it takes longer times for the e2 and p2 to reach up to the detectable flux levels. However, whether this delay is due to the times that needed for the evolving shock to be strengthened or due to particle transport effects are not resolved.

%Figure 10 shows that these events either have a slow rise in the e2 and p2 channels or  where the real first arriving particle onset have been masked by the background level, which may be well below the background level.
%All these events have a relatively strong anisotropy value, indicating directly injections at shock acceleration occured for the 65 keV electrons, but not for the e2 and p2 particles.
%Due to the soft spectrum, the e2 or p2 flux enhancements are too small to be detected at the e1 SPR time. 

\begin{acknowledgments}
The authors would like to thank the support of STEREO, SOHO, WIND and ACE teams. 
The STEREO SECCHI data are produced by a consortium of RAL (UK), NRL (USA), LMSAL 
(USA), GSFC (USA), MPS (Germany), CSL (Belgium), IOTA (France), and IAS (France). 
The SOHO LASCO data are produced by a consortium of the Naval Research 
Laboratory (USA), Max-Planck-Institut f$\ddot{u}$r Aeronomie (Germany), Laboratoire 
d'Astronomie (France), and the University of Birmingham (UK).
SOHO Electron Proton and Helium Instrument (EPHIN) data were obtained from:
\url{ http://www2.physik.uni-kiel.de/SOHO/phpeph/EPHIN.htm};
SOHO Energetic and Relativistic Nuclei and Electron instrument (ERNE) data were obtained from:
\url{http://www.srl.utu.fi/erne_data/datafinder/df.shtml
};
STEREO High Energy Telescope (HET) data were obtained from:
\url{http://www.srl.caltech.edu/STEREO/Public/HET_public.html};
STEREO High Energy Telescope (LET) data were obtained from:
\url{http://www.srl.caltech.edu/STEREO/Public/LET_public.html};
STEREO Solar Electron Proton Telescope data (SEPT) were obtained from:
\url{http://www2.physik.uni-kiel.de/STEREO/index.php?doc=data};
 and Wind/3DP and ACE/EPAM proton and electron data were obtained from
\url{http://cdaweb.gsfc.nasa.gov/istp_public/}.
This work was supported by NASA LWS TR\&T program NNX15AB70G.
PM was partially supported by NASA grant NNX15AB77G and NSF grant AGS-1358274.
\end{acknowledgments}

% If you use BiBTeX for your References, please produce your .bbl
% file and copy the contents into your paper here.
%
% Follow these steps:
% 1. Run LaTeX on your LaTeX file.
%
% 2. Run BiBTeX on your LaTeX file.
%
% 3. Open the new .bbl file containing the reference list and
%   copy all the contents into your LaTeX file here.
%
% 4. Comment out the old \bibliographystyle and \bibliography commands.
%
% 5. Run LaTeX on your new file before submitting.
%
% AGU does not want a .bib or a .bbl file, but asks that you
% copy in the contents of your .bbl file here.

%\bibliography{agu3_update}
%\bibliographystyle{agu08}

%% ------------------------------------------------------------------------ 
%%
%
%  END ARTICLE
%
%% ------------------------------------------------------------------------ 
%%

\end{article}

\begin{landscape}
\begin{table}
\caption{SEP Electrons and Protons Solar Release Times and Associated Solar Eruptive Signatures }

\label{Table1}
\resizebox{1.3\textwidth}{!}{%
\begin{tabular}{cccccccccccccccc}
\tableline
$\#$    &  Date        &\multicolumn{2}{c} {EPHIN SPR} & \multicolumn{3}{c} {ERNE SPR} & Type& \multicolumn{2}{c} {Type II} & Spd & Loc&Ht& SC& CA\tablenotemark{d} \\
       &              & e1& e2 & p1 & p2 & p3 & III& mII & DH &  &  &  &  &  &\\
       &              & \multicolumn{2}{c} {UT} &\multicolumn{3}{c} {UT} & UT &UT& UT &   $km/s$  & &Rs &  &  $\circ$ &  \\                                 
 \tableline
  1 \tablenotemark{a}  &  2006/12/13  &    02:30 $\pm$ 3  & 02:39 $_{ -12}^{ +3}$ & 02:36 $\pm$ 5  & 02:36 $\pm$ 5  &  ---:---       & 02:26 & 02:26  & 02:45 & 1774 & S06W23    & 2.7  & S &     16 \\     
  2\tablenotemark{c}   &  2006/12/14  &   22:17 $\pm$ 3  & 22:24 $\pm$ 11 & 22:23 $\pm$ 5  & 22:24 $\pm$ 5  &  ---:---         & 22:10 & 22:09  & 22:30 & 1042   & S06W46    & 2.4  & S &   11 \\     
  3                    &  2010/08/14  &   10:05 $\pm$ 3  & ---:---        & ---:---        & 10:12 $\pm$ 5  &  10:15 $\pm$  5  & 09:56 & 09:52  & 10:00 & 1205 & N17W52    & 2.9  & S &   30 \\     
  4                    &  2011/03/07  &   20:18 $\pm$ 5  & ---:---        & ---:---        & 20:27 $\pm$ 5  &  20:30 $\pm$ 13  & 19:52 & 19:54  & 20:00 & 2125 & N31W53    & 8.7  & S &    2 \\     
  5                    &  2011/06/07  &   06:41 $\pm$ 12 & 06:45 $\pm$  7 & 06:42 $\pm$ 8  & 06:40 $\pm$ 12 &  ---:---         & 06:26 & 06:25  & 06:45 & 1255 & S21W54    & 3.8  & S &    1 \\     
  6                    &  2011/08/04  &   04:17 $\pm$ 3  & 04:31 $\pm$  10 & 04:23 $\pm$ 5  & 04:18 $\pm$ 5  &  04:22 $\pm$ 14 & 03:52 & 03:54  & 04:15 & 1315 & N19W36    & 6.0  & S &  13 \\       
  7\tablenotemark{a}   &  2011/08/09  &   08:04 $\pm$ 3  & 08:12$_{ -12}^{ +3}$ & 08:01 $\pm$ 5  & 08:00 $\pm$ 5  &  07:59 $\pm$  5  & 08:02 & 08:01  & 08:20 & 1610 & N17W69    & 2.1  & S &    2 \\       
  8\tablenotemark{b}   &  2011/11/26  &   07:20 $\pm$ 3  & 07:29 $\pm$  7 & ---:---        & 07:45 $\pm$ 10  &  07:46 $\pm$ 10 & 07:10 & ---:---& 07:15 &  933 & N17W49    & 3.5  & S &   6 \\       
  9                    &  2012/01/27  &   18:22 $\pm$ 3  & 18:27 $\pm$  3 & 18:30 $\pm$ 5  & 18:28 $\pm$ 5  &  18:45 $\pm$ 12  & 18:16 & 18:10  & 18:30 & 2508 & N27W71    & 3.2  & S &   -9 \\       
  10                   &  2012/03/13  &   17:31 $\pm$ 3  & 17:35 $\pm$  3 & 17:32 $\pm$ 5  & 17:32 $\pm$ 5  &  17:29 $\pm$  9  & 17:17 & 17:15  & 17:35 & 1884 & N19W66    & 3.3  & S &  -27 \\       
  11                   &  2012/05/17  &   01:34 $\pm$ 3  & 01:40 $\pm$  3 & 01:37 $\pm$ 5  & 01:31 $\pm$ 5  &  01:33 $\pm$  5  & 01:32 & 01:31  & 01:40 & 1582 & N11W76    & 2.3  & S &   -8  \\       
  12\tablenotemark{b}  &  2012/07/12  &   16:51 $\pm$ 5  & ---:---        & ---:---         & 16:58 $\pm$ 10  &  16:59 $\pm$  10  & 16:31 & 16:25  & 16:45 &  885 & S15W01    & 2.9  & S & 54  \\       
  13                   &  2012/07/17  &   14:43 $\pm$ 5  & ---:---        & ---:---        & 14:43 $\pm$ 10  & 14:51 $\pm$ 10  & 14:01 & ---:---& 14:40 &  958 & S28W75    & 4.4  & S &  -23  \\       
  14\tablenotemark{c}  &  2012/07/19  &   ---:---        & ---:---        & 06:06 $\pm$ 15 & 05:53 $\pm$ 11 &  ---:---         & 05:25 & 05:24  & 05:30 & 1631 & S13W88    & 9.1  & S &  -33  \\       
  15                   &  2013/05/22  &   13:29 $\pm$ 6  & 13:32 $\pm$  3 & 13:30 $\pm$ 5  & 13:27 $\pm$ 5  &  13:26 $\pm$  6  & 13:10 & ---:---& 13:10 & 1466 & N13W75    & 6.0  & S &  -34  \\       
  16\tablenotemark{c}  &  2014/01/07  &   18:38 $\pm$ 9  & 18:47 $\pm$  5 & 18:43 $\pm$ 10 & 18:46 $\pm$ 14 &  ---:---         & 18:04 & 18:17  & 18:27 & 1830 & S19W29    & 8.6  & S &   33  \\     
  17                   &  2014/02/20  &   07:50 $\pm$ 3  & 07:54 $\pm$  3 & 07:51 $\pm$ 5  & 07:49 $\pm$ 5  &  07:46 $\pm$  15  & 07:46 & 07:45  & 08:06 & 1040 & S15W73   & 2.8  & S &  -24  \\     
  S1                   &  2011/03/21  &  02:25  $\pm$ 5  & 02:31  $\pm$ 5  & 02:32  $\pm$ 5 & 02:31 $\pm$ 5  &  02:33 $\pm$  5  & 02:21 & ---:---& 02:30 & 1341 & N26W41   & 2.6  & A &   14 \\     
  S2                   &  2011/09/22  &  10:47  $\pm$ 5  & 10:55  $\pm$ 5  & 10:53  $\pm$ 5 & 10:54 $\pm$ 5  &  11:05 $\pm$  5  & 10:40 & 10:39  & 11:05 & 1905 & N09W07   & 2.7  & B &   46 \\     
  S3\tablenotemark{c}  &  2012/03/07  &  00:42  $\pm$ 9  & 00:44  $\pm$ 6  & 00:46  $\pm$ 5 & 00:51 $\pm$ 5  &  01:14 $\pm$  22 & 00:18 & 00:17  & 00:36 & 2684 & N22W105  & 7.0  & B &  -41 \\     
  S4                   &  2012/05/26  &  20:57  $\pm$ 5  & 21:02  $\pm$ 6  & 21:04  $\pm$ 7 & 20:59 $\pm$ 6  &  21:02 $\pm$  5  & 20:46 & 20:47  & 20:50 & 1966 & N11W11   & 2.4  & A &   48 \\     
  S5\tablenotemark{c}  &  2012/07/23  &  02:25  $\pm$ 5  & 02:53  $\pm$ 16 & 02:50  $\pm$ 6 & 02:49 $\pm$ 5  &  02:48 $\pm$  6  & 02:14 & ---:---& 02:30 & 2003 & N05W15   & 3.0  & A &   32 \\     
  S6                   &  2012/08/31  &  20:03  $\pm$ 5  & 20:05  $\pm$ 5  & 20:18  $\pm$ 5 & 20:17 $\pm$ 5  &  20:21 $\pm$  5  & 19:45 & 19:42  & 20:00 & 1442 & S19W73   & 3.0  & B &    3 \\    
  S7\tablenotemark{c}  &  2013/03/15  &  07:03  $\pm$ 10 & ---:---         & ---:---        & ---:---        &  07:27 $\pm$  18 & 06:39 & ---:---& 07:00 & 980 & N11W128  & 3.9  & B &  -58 \\     
  S8                   &  2013/04/11  &  07:19  $\pm$ 5  & 07:21  $\pm$ 5  & 07:27  $\pm$ 5 & 07:24 $\pm$ 5  &  07:22 $\pm$  5  & 07:03 & 07:02  & 07:10 &  861 & N09W129  & 1.7  & B &  -54 \\     
  S9                   &  2013/05/13  &  02:20  $\pm$ 5  & 02:24  $\pm$ 6  & 02:37  $\pm$ 9 & 02:41 $\pm$ 7  &  02:47 $\pm$ 10  & 02:08 & 02:10  & 02:20 & 1366 & N12W77   & 2.2  & B &  -16 \\     
  S10                  &  2013/06/21  &  03:04  $\pm$ 10  & 03:16 $\pm$ 12 & ---:---        & ---:---        &  03:45 $\pm$ 15  & 02:51 & ---:---& 03:36 & 1900 & S16W66   & 2.8  & B &  -21 \\     
  S11                  &  2014/02/25  &  01:05  $\pm$ 5  & 01:09  $\pm$ 5  & 01:05  $\pm$ 5 & 01:12 $\pm$ 5  &  01:22 $\pm$  5  & 00:46 & 00:56  & 01:02 & 2147 & S12W78   & 3.7  & B &  -35 \\     
 \tableline
\end{tabular}
}
\tablenotetext{a}{X-ray contamination electron event}
\tablenotetext{b}{Data from WIND/EPACT protons being used when SOHO's roll angle is 180\deg.}
\tablenotetext{c}{High background flux level event}
\tablenotetext{d}{With a large uncertainty of 20--30 \deg}
\end{table}
\end{landscape}

\begin{table}
\caption{Electron Anisotropy and Velocity Dispersion Analysis Results for Protons}
\label{Table3}
\begin{tabular}{cccccccc}
\tableline
$\#$  &    Date   & Aniso A &\multicolumn{2}{c}{ Electron } &   \multicolumn{3}{c} {Proton VDA}             \\
      &           &     &   SPR\_e1 & SPR\_e2 &                  SPR\_p           &   $L_{path}$   &  energy (low-high)  \\
      &           &          & (UT)  & (UT) &                 (UT)           &    (AU)           &      (MeV)    \\      
\tableline
  1   &  2006/12/13   & 1.07   &02:30 &  02:39 &    02:31  $\pm$  1.76   & 1.54  $\pm$  0.08 &   57.4   90.5 \\    
  2   &  2006/12/14   & 1.17   &22:17 &  22:24 &    HBG                  & ---              &   ---    --- \\ 
  3   &  2010/08/14   &-0.90   &10:05 &  --:-- &    10:13  $\pm$  3.03   & 1.27  $\pm$  0.09 &   15.4  57.4  \\ 
  4   &  2011/03/07   &-1.42   &20:18 &  --:-- &    20:17  $\pm$  4.0    & 1.61  $\pm$  0.10 &   ---    ---\\ 
  5   &  2011/06/07   &-1.24   &06:41 &  06:45 &    06:47  $\pm$  4.0    & 1.35  $\pm$  0.10 &   25.3  50.4  \\ 
  6   &  2011/08/04   &-1.55   &04:17 &  04:31 &    04:36  $\pm$  6.35   & 1.19  $\pm$  0.17 &   15.4  45.6  \\ 
  7   &  2011/08/09   & 1.41   &08:04 &  08:12 &    08:03  $\pm$  6.06   & 1.18  $\pm$  0.24 &   57.4  72.0  \\ 
  8   &  2011/11/26   &-2.09   &07:20 &  07:29 &    07:35  $\pm$  4.0    & 1.30  $\pm$  0.10 &   25.3  50.4  \\ 
  9   &  2012/01/27   &-0.27   &18:22 &  18:27 &    18:27  $\pm$  12.71  & 2.51  $\pm$  0.33 &   15.4  57.4  \\ 
  10  &  2012/03/13   &-1.96   &17:31 &  17:35 &    17:34  $\pm$  2.82   & 1.24  $\pm$  0.09 &   15.4  90.5  \\ 
  11  &  2012/05/17   &-2.23   &01:34 &  01:40 &    IC                   & ---              &   ---    ---\\ 
  12  &  2012/07/12   & 2.13   &16:57 &  --:-- &    16:56  $\pm$  4.0    & 1.30  $\pm$  0.10 &   25.3  50.4  \\ 
  13  &  2012/07/17   &-1.10   &14:43 &  --:-- &    14:44  $\pm$  0.66   & 1.19  $\pm$  0.02 &   15.4  45.6  \\ 
  14  &  2012/07/19   &-0.44   &--:-- &  --:-- &    HBG                  & ---              &   ---    ---\\ 
  15  &  2013/05/22   &-1.07   &13:29 &  13:32 &    13:30  $\pm$   4.0   & 1.36  $\pm$  0.10 &   0.0  50.4  \\ 
  16  &  2014/01/07   & 0.65   &18:38 &  18:47 &    HBG                  & ---              &   ---    ---\\ 
  17  &  2014/02/20   & 1.94   &07:50 &  07:54 &   07:51  $\pm$   3.86   & 1.20  $\pm$  0.14 &   23.3  57.4 \\ 
  S1  &  2011/03/21   &-1.37   &02:25 &  02:31&    02:29  $\pm$   1.63   & 1.25  $\pm$ 0.04  &  11.0  50.0   \\ 
  S2  &  2011/09/22   &  2.24  &10:47 &  10:55 &   10:47  $\pm$   4.45   & 1.58  $\pm$ 0.10  &  11.0  38.0   \\ 
  S3  &  2012/03/07   & -1.08  &00:42 &  00:44 &   00:42  $\pm$   9.79   & 1.78  $\pm$ 0.23  &  11.0  50.0   \\ 
  S4  &  2012/05/26   &  1.98  &20:57 &  21:02 &   20:58  $\pm$   3.74   & 1.42  $\pm$ 0.09  &  11.0  50.0   \\ 
  S5  &  2012/07/23   &  1.28  &02:25 &  02:53 &   02:56  $\pm$   1.34   & 1.26  $\pm$ 0.05  &  38.0  50.0   \\ 
  S6  &  2012/08/31   &  2.97  &20:03 &  20:05 &   20:16  $\pm$   0.86   & 1.56  $\pm$ 0.02  &  18.1  38.0   \\ 
  S7  &  2013/03/15   &  1.44  &07:03 &  --:-- &   NA                    & ---              &  ---  ---   \\ 
  S8  &  2013/04/11   &  2.52  &07:19 &  07:21 &   07:22  $\pm$   0.68   & 1.50  $\pm$ 0.02  &  11.0  50.0   \\ 
  S9  &  2013/05/13   & -0.92  &02:20 &  02:24 &   02:33  $\pm$   0.04   & 1.65  $\pm$ 0.00  &  11.0  50.0   \\ 
 S10  &  2013/06/21   &  1.76  &03:04 &  03:16 &   03:34  $\pm$   1.39   & 1.54  $\pm$ 0.03  &  11.0  25.1   \\ 
 S11  &  2014/02/25   &  NA    &01:05 &  01:09 &   01:09  $\pm$   1.87   & 1.40  $\pm$ 0.05  &  11.0  50.0   \\ 
\tableline
\end{tabular}
%\tablenotetext{a}{Columns 1 - 4: CME1 first appearance time on C2, FR-fit radial speed, face-on half width and propagation direction to COR2 images}
\end{table}

\begin{table}
\caption{Energy channels used in the SOHO/ERNE velocity dispersion analysis}
\label{Table2}
\begin{tabular}{ccccc}
\tableline
      & Channel & Energy range (MeV) & Average energy (MeV) &  Inverse speed ($\beta^{-1}$) \\
      &   0    &     1.58-1.78     &       1.68         &         16.7     \\ 
LED   &   1    &     1.78-2.16     &       1.97         &         15.5     \\ 
      &   2    &     2.16-2.66     &       2.41         &         14.0     \\ 
      &   3    &     2.66-3.29     &       2.98         &         12.6     \\ 
      &   4    &     3.29-4.10     &       3.70         &         11.3     \\ 
      &   5    &     4.10-5.12     &       4.71         &         10.0     \\ 
      &   6    &     5.12-6.42     &       5.72         &         9.10     \\ 
      &   7    &     6.42-8.06     &       7.15         &         8.15     \\ 
      &   8    &     8.06-10.1     &       9.09         &         7.24     \\ 
      &   9    &     10.1-12.7     &       11.4         &         6.47     \\                                                   
\tableline                                                         
HED    &    10   &     13.8-16.9     &       15.4        &        5.59     \\ 
       &    11   &     16.9-22.4     &       18.9        &        5.06     \\ 
       &    12   &     20.8-28.0     &       23.3        &        4.57     \\ 
       &    13   &     25.9-32.2     &       29.1        &        4.11     \\ 
       &    14   &     32.2-40.5     &       36.4        &        3.69     \\ 
       &    15   &     40.5-53.5     &       45.6        &        3.32     \\ 
       &    16   &     50.8-67.3     &       57.4        &        2.99     \\ 
       &    17   &     63.8-80.2     &       72.0        &        2.70     \\ 
       &    18   &     80.2-101      &       90.5        &        2.44     \\ 
       &    19   &      101-131      &       108         &        2.26     \\   
\tableline
\end{tabular}
%\tablenotetext{a}{Columns 1 - 4: CME1 first appearance time on C2, FR-fit radial speed, face-on half width and propagation direction to COR2 images}
\end{table}

\end{document}